\documentclass[final,3p,times,nopreprintline]{elsarticle}

\usepackage{amssymb}
\usepackage{amsmath}
\usepackage{booktabs}
\usepackage{subfig}

\usepackage{graphicx}
\usepackage{algorithm}
\usepackage{algorithmic}
\usepackage{multirow}

\begin{document}

\begin{frontmatter}

\title{Deep Reinforcement Learning with Massively Parallel Simulation for Low-Thrust Multi-Revolution Rendezvous}

\author[label1]{Chenglong Wu} 
\author[label1]{Yi Zhou} 
\author[label1]{Bin Cheng} 
\author[label2]{Nan Zhang} 
\author[label1]{Hexi Baoyin\corref{cor1}} 
\ead{baoyin@tsinghua.edu.cn}

\cortext[cor1]{Corresponding author}

\affiliation[label1]{organization={Tsinghua University}, 
            addressline={30 Shuangqing Rd, Haidian District}, 
            city={Beijing},
            postcode={100084}, 
            state={Beijing},
            country={China}}
\affiliation[label2]{organization={Beihang University}, 
            addressline={37 Xueyuan Rd, Haidian District}, 
            city={Beijing},
            postcode={100191}, 
            state={Beijing},
            country={China}}

\begin{abstract}
The growing demand for on-orbit servicing and constellation management in low Earth orbit requires autonomous and reliable low-thrust rendezvous. Traditional numerical methods struggle with the long-horizon nature of multi-revolution maneuvers under {perturbations}, while standard deep reinforcement learning (DRL) methods are limited by low simulation throughput and poor sample efficiency. To address these challenges, this paper proposes a massively parallel DRL guidance framework. A fully tensorized, GPU-resident orbital propagator is developed to integrate {$J_2$-perturbed orbital dynamics} and policy optimization within a zero-copy pipeline, achieving up to a 67-fold increase in simulation throughput over conventional CPU baselines. In addition, the rendezvous problem is formulated in modified equinoctial elements and trained with a progressive curriculum learning strategy to improve convergence and generalization across increasingly challenging rendezvous scenarios. Simulations show that the proposed framework can learn robust rendezvous trajectories for complex multi-revolution rendezvous over a broad range of orbital configurations. Additional evaluations under {a high-fidelity model and} eclipse-induced thrust interruptions further demonstrate the robustness of the learned policy under practical mission constraints. Moreover, the trajectories generated by the proposed framework can be used as effective initial guesses for optimization-based trajectory solvers, improving the convergence of high-precision trajectory optimization. The method offers a scalable approach for real-time low-thrust guidance and trajectory optimization.
\end{abstract}


\end{frontmatter}

\section{Introduction}

The rapid deployment of mega-constellations, together with the growing demand for On-Orbit Servicing and Active Debris Removal, has rendered the space environment---particularly low Earth orbit (LEO) at altitudes of approximately 500 km---increasingly congested. In such highly dynamic orbital regimes, satellites increasingly need to achieve autonomous, rapid, and reliable rendezvous maneuvering. Meanwhile, to maximize payload mass fractions and extend operational lifetimes, modern satellites increasingly adopt electric propulsion. Although electric propulsion offers extremely high specific impulse, its thrust acceleration is minute (typically $10^{-4}$--$10^{-3}\,\mathrm{m/s^2}$), which forces low-thrust rendezvous to operate continuously or near-continuously over long orbital arcs and turns trajectory design into a long-horizon optimal control problem with strongly nonlinear dynamics and environmental perturbations~\cite{grahamMinimumTimeTrajectoryOptimization2015}.

Traditional low-thrust trajectory optimization methods are mainly categorized into indirect and direct methods~\cite{bettsSurveyNumericalMethods1998,colasurdoOptimalFinitethrustSpacecraft1992,enrightOptimalFinitethrustSpacecraft1991,zavoliIndirectOptimizationFiniteThrust2015}. Indirect methods leverage Pontryagin's Maximum Principle (PMP) to derive first-order necessary conditions and solve the resulting two-point boundary value problem. While they can be highly accurate with low decision-variable dimension, they typically suffer from narrow convergence domains and strong sensitivity to initial costate guesses, whose physical interpretation is often non-intuitive. Direct methods discretize the continuous problem into a Nonlinear Programming (NLP) problem (e.g., via collocation)~\cite{enrightOptimalFinitethrustSpacecraft1991}. They avoid explicit costate initialization and usually admit broader convergence, but long-horizon low-thrust problems lead to extremely high-dimensional NLPs, heavy computational burdens for onboard use, and accuracy limitations tied to the mesh resolution, making it difficult to tightly satisfy continuous-time optimality conditions~\cite{raoSurveyNumericalMethods2009}. As a result, both classes of approaches tend to rely on terrestrial offline computation and are difficult to meet real-time, autonomous, closed-loop guidance requirements in congested and perturbation-rich LEO environments.

To reduce online computational complexity, sequential convex programming has been explored for real-time trajectory optimization by iteratively convexifying the non-convex dynamics/constraints around a reference trajectory and solving a sequence of convex subproblems with polynomial-time guarantees~\cite{szmukSuccessiveConvexification6DoF2018}. Sequential convex programming has shown success in powered descent and reusable launch vehicle guidance~\cite{blackmoreAutonomousPrecisionLanding2016}. However, multi-revolution low-thrust rendezvous typically spans long arcs under complex perturbations; in such strongly coupled nonlinear evolution, local linearization may introduce unacceptable truncation errors, leading to convexification breakdown or divergence. In parallel, supervised learning has been used to imitate offline optimal solutions~\cite{sanchez-sanchezRealTimeOptimalControl2018}, but its performance and robustness are fundamentally limited by dataset coverage and can degrade significantly under out-of-distribution disturbances.

Reinforcement Learning (RL) offers an alternative by learning a closed-loop state-to-control policy through interaction with the dynamical environment, thereby avoiding explicit solution of PMP conditions or large-scale NLPs~\cite{izzoSurveyArtificialIntelligence2019,bertsekasFeaturebasedAggregationDeep2018,suttonReinforcementLearningIntroduction1998}. After computationally intensive ground training, the onboard neural policy can generate thrust commands via fast forward propagation, enabling an ``offline training, online execution'' paradigm for real-time guidance~\cite{hovellDeepReinforcementLearning2021}. Deep reinforcement learning (DRL) has demonstrated promising performance in a variety of aerospace tasks, including terminal guidance and powered descent under uncertainties~\cite{tipaldiReinforcementLearningSpacecraft2022,gaudetAdaptiveGuidanceIntegrated2020,jiangIntegratedGuidanceMars2019,scorsoglioImageBasedDeepReinforcement2022,ponnarasanReinforcementLearningAutonomous2025}, and has also been explored for long-horizon deep-space and interplanetary transfers~\cite{zavoliReinforcementLearningRobust2021,miller2019interplanetary}. Nevertheless, multi-revolution low-thrust rendezvous remains particularly challenging due to non-monotonic orbital phase evolution and severe long-term credit assignment~\cite{smith2021propulsionless}, as well as the highly non-convex trade-off between fuel efficiency and terminal precision~\cite{tang2018fuel}. Moreover, high-fidelity orbital simulations require accurate numerical integration and coupled constraint handling, making environment rollouts far more expensive than standard RL benchmarks; combined with DRL's large sample complexity, conventional serial CPU-based training becomes computationally prohibitive.

To tackle these challenges, this paper proposes a soft actor-critic algorithm based guidance framework with massive GPU parallel acceleration and progressive curriculum learning (CL). Computationally, we build a fully tensorized orbital dynamics propagator in JAX \cite{jax2018github}, transforming the traditional serial CPU integration into batchable, vectorized state propagation. Orbital simulation, policy inference, gradient backpropagation, and parameter updates are integrated into a unified GPU-resident computational graph, avoiding repeated CPU--GPU data transfers and reducing the overhead of serial environment rollouts, thereby improving sample throughput and training efficiency. Algorithmically, SAC's maximum-entropy formulation and off-policy learning improve exploration robustness and sample efficiency in long-horizon continuous thrust control~\cite{tipaldiReinforcementLearningSpacecraft2022,suttonReinforcementLearningIntroduction1998}.Rather than a strictly time-optimal transfer, we formulate a multi-objective optimization problem. Guided by penalties on both fuel consumption and per-step distance, the agent is incentivized to find a practical trade-off between fuel efficiency and rapid rendezvous.To mitigate sparse rewards and long-term credit assignment, a progressive curriculum expands phase/altitude offsets from easy to hard, improving convergence and generalization for large-phase-angle multi-revolution rendezvous~\cite{smith2021propulsionless,tang2018fuel}.

The rest of this paper is organized as follows. Section~\ref{sec2} formulates the low-thrust rendezvous problem, including coordinate systems, $J_2$-perturbed dynamics, and the optimal control objective. Section~\ref{sec3} presents the proposed massively parallel DRL framework, including the JAX-based architecture, Markov Decision Process (MDP) formulation, SAC algorithm, and curriculum strategy. {Section~\ref{sec:results} evaluates robustness under the high-fidelity model, including tests with eclipse-induced thrust interruptions and navigation errors. It then presents optimality comparisons, curriculum-learning ablations, and computational-efficiency analyses under the nominal $J_2$ model.} Section~\ref{sec:conclusion} concludes the paper.

\section{Problem Formulation}
\label{sec2}

\subsection{Coordinate Systems}
\label{2-subsec1}

The state of the satellite is described using modified equinoctial elements (MEEs) for the low-thrust problem.  The MEE state vector $\boldsymbol{x} = [p, f, g, h, k, L]^T$ is chosen for its nonsingular properties, particularly for the near-circular orbits common in LEO rendezvous. Its components are related to the Classical Orbital Elements (COEs) via:
\begin{equation}
\label{eq:mee_def}
\left\{
\begin{aligned}
    p &= a(1 - e^2) \\
    f &= e \cos(\omega + \Omega) \\
    g &= e \sin(\omega + \Omega) \\
    h &= \tan(i/2) \cos\Omega \\
    k &= \tan(i/2) \sin\Omega \\
    L &= \Omega + \omega + \nu
\end{aligned}
\right.
\end{equation}
while the control thrust vector $\boldsymbol{u}$ is defined in the Local-Velocity/Local-Horizon (LVLH) frame, the basis vectors $\{\boldsymbol{e}_r, \boldsymbol{e}_t, \boldsymbol{e}_n\}$ are aligned with the satellite's instantaneous radial, transverse, and normal directions, respectively, and are defined by:
\begin{equation}
    \boldsymbol{e}_r = \frac{\boldsymbol{r}}{\|\boldsymbol{r}\|}, \quad 
    \boldsymbol{e}_n = \frac{\boldsymbol{r} \times \boldsymbol{v}}{\|\boldsymbol{r} \times \boldsymbol{v}\|}, \quad 
    \boldsymbol{e}_t = \boldsymbol{e}_n \times \boldsymbol{e}_r
\end{equation}

From the perspective of RL, this transforms the state space into a much smoother landscape, which is more effective in solving complex, multi-revolution rendezvous problems.

\subsection{Low Thrust Dynamic Model with $J_2$ Perturbation}
\label{2-subsec2}
The dynamic model of the satellite is governed by Gauss's Variational Equations (GVE). In the context of MEE, the state dynamics can be expressed in a compact matrix form. Let the state vector be $\boldsymbol{x} = [p, f, g, h, k, L]^T$ and the mass of the satellite  be $m$. The differential equations of motion are given by the following:

\begin{equation}
\label{eq:gve_matrix}
    \dot{\boldsymbol{x}} = \boldsymbol{B}(\boldsymbol{x}) \cdot \left( \boldsymbol{a}_T + \boldsymbol{a}_{J_2} \right) + \boldsymbol{P}
\end{equation}

\noindent where $\boldsymbol{a}_T$ represents the control thrust acceleration, $\boldsymbol{a}_{J_2}$ denotes the gravitational perturbation acceleration due to the oblateness of the Earth, and $\boldsymbol{P} = [0, 0, 0, 0, 0, \sqrt{\mu p} (w/p)^2]^T$ presents the natural angular motion of the true longitude. The matrix $\boldsymbol{B}(\boldsymbol{x}) \in \mathbb{R}^{6 \times 3}$ maps the accelerations in the LVLH frame to the variations of the MEE state:

\begin{equation}
\label{eq:B_matrix}
    \boldsymbol{B}(\boldsymbol{x}) = \sqrt{\frac{p}{\mu}}
    \begin{bmatrix}
        0 & \frac{2p}{w} & 0 \\
        \sin L & \frac{(w+1)\cos L + f}{w} & -\frac{g(h \sin L - k \cos L)}{w} \\
        -\cos L & \frac{(w+1)\sin L + g}{w} & \frac{f(h \sin L - k \cos L)}{w} \\
        0 & 0 & \frac{s^2 \cos L}{2w} \\
        0 & 0 & \frac{s^2 \sin L}{2w} \\
        0 & 0 & \frac{h \sin L - k \cos L}{w}
    \end{bmatrix}
\end{equation}

\noindent Here, $\mu$ is the gravitational parameter of the Earth. The auxiliary variables $w$ and $s^2$ are defined as:
\begin{equation}
    w = 1 + f \cos L + g \sin L, \quad s^2 = 1 + h^2 + k^2
\end{equation}

The total perturbation acceleration vector is resolved in the LVLH frame. The control acceleration $\boldsymbol{a}_T$ is generated by the on-board propulsion system. Let the control action vector be $\boldsymbol{u} = [u_r, u_t, u_n]^T$, representing the thrust direction and magnitude normalized by the maximum thrust $T_{max}$. The control acceleration is explicitly defined as:
\begin{equation}
    \boldsymbol{a}_T = \frac{T_{max}}{m} \boldsymbol{u}, \quad \text{s.t.} \quad \|\boldsymbol{u}\|_\infty \le 1
\end{equation}

To ensure physical fidelity, the dominant perturbation $J_2$ is included. The acceleration of perturbation components in the LVLH frame, denoted  $\boldsymbol{a}_{J_2} = [a_{r,J2}, a_{t,J2}, a_{n,J2}]^T$, are calculated as follows \cite{battin1999introduction}:
\begin{equation}
\label{eq:j2_pert}
\begin{aligned}
    a_{r,J2} &= -\frac{3 \mu J_2 R_e^2}{2 r^4} \left( 1 - 12 \frac{(h \sin L - k \cos L)^2}{s^4} \right) \\
    a_{t,J2} &= -\frac{12 \mu J_2 R_e^2}{r^4} \left( \frac{(h \sin L - k \cos L)(h \cos L + k \sin L)}{s^4} \right) \\
    a_{n,J2} &= -\frac{6 \mu J_2 R_e^2}{r^4} \left( \frac{(1-h^2-k^2)(h \sin L - k \cos L)}{s^4} \right)
\end{aligned}
\end{equation}
where $R_e$ is the mean equatorial radius of the Earth and $r = p/w$ is the instantaneous orbital radius. 

\subsection{Low-Thrust Rendezvous Optimization Problem}
\label{2-subsec3}
Based on the established coordinate system and dynamic model, the low-thrust rendezvous mission is formulated as a final-time-free optimal control problem. 
Let $\boldsymbol{x}(t) \in \mathbb{R}^6$ denote the MEE state of the chaser satellite and $\boldsymbol{x}_d(t)$ denote the state of the target satellite. Since the target is assumed to be passive, $\boldsymbol{x}_d(t)$ propagates naturally under $J_2$ perturbation. The rendezvous condition requires that the position and velocity of the chaser match those of the target at $t_f$, which is described using MEEs as follows:
\begin{equation}
    \boldsymbol{x}(t_f) = \boldsymbol{x}_d(t_f)
\end{equation}
To mathematically quantify the performance, we define a composite cost function in Lagrange form. The objective is to find the optimal control policy $\boldsymbol{u}^*(t)$ and the terminal time $t_f^*$ that minimize the objective:
\begin{equation}
\label{eq:cost_function}
    J = \int_{t_0}^{t_f} \left( 1 + \lambda \|\boldsymbol{u}(t)\|^2 \right) dt
\end{equation}
where $\lambda > 0$ is a hyperparameter that regulates the relative importance of fuel saving versus time efficiency. The complete low-thrust rendezvous problem can be summarized as follows:
\begin{equation}
\label{eq:ocp_summary}
\begin{aligned}
    \min_{\boldsymbol{u}(\cdot), t_f} \quad & J = \int_{t_0}^{t_f} (1 + \lambda \|\boldsymbol{u}\|^2) dt \\
    \textrm{s.t.} \quad & \dot{\boldsymbol{x}}(t) = \boldsymbol{B}(\boldsymbol{x}) \cdot \left( \frac{T_{max}}{m} \boldsymbol{u} + \boldsymbol{a}_{J_2} \right) + \boldsymbol{P} \\
    & \dot{\boldsymbol{x}}_d(t) = \boldsymbol{B}(\boldsymbol{x}_d) \cdot \boldsymbol{a}_{J_2,d} + \boldsymbol{P}_d \\
    & \|\mathbf u\|_\infty \le 1, \quad \forall t \in [t_0, t_f] \\
    & \boldsymbol{x}(t_0) = \boldsymbol{x}_{0}, \quad \boldsymbol{x}_d(t_0) = \boldsymbol{x}_{d,0} \\
    & \boldsymbol{x}(t_f) - \boldsymbol{x}_d(t_f) = \boldsymbol{0}
\end{aligned}
\end{equation}

It is worth noting that while the theoretical objective in Eq.~\eqref{eq:cost_function} utilizes a constant time penalty to minimize transfer duration, our MDP formulation employs a distance-based penalty. This acts as a dense reward shaping technique, which provides stronger gradient signals to accelerate convergence while implicitly maintaining the time-efficiency objective by driving the agent toward the target as quickly as possible.

\section{Methodology}
\label{sec3}
\subsection{Overview of the Training Framework}
\label{3-overview}
\begin{figure}[htbp]
    \centering
    \includegraphics[width=0.7\linewidth]{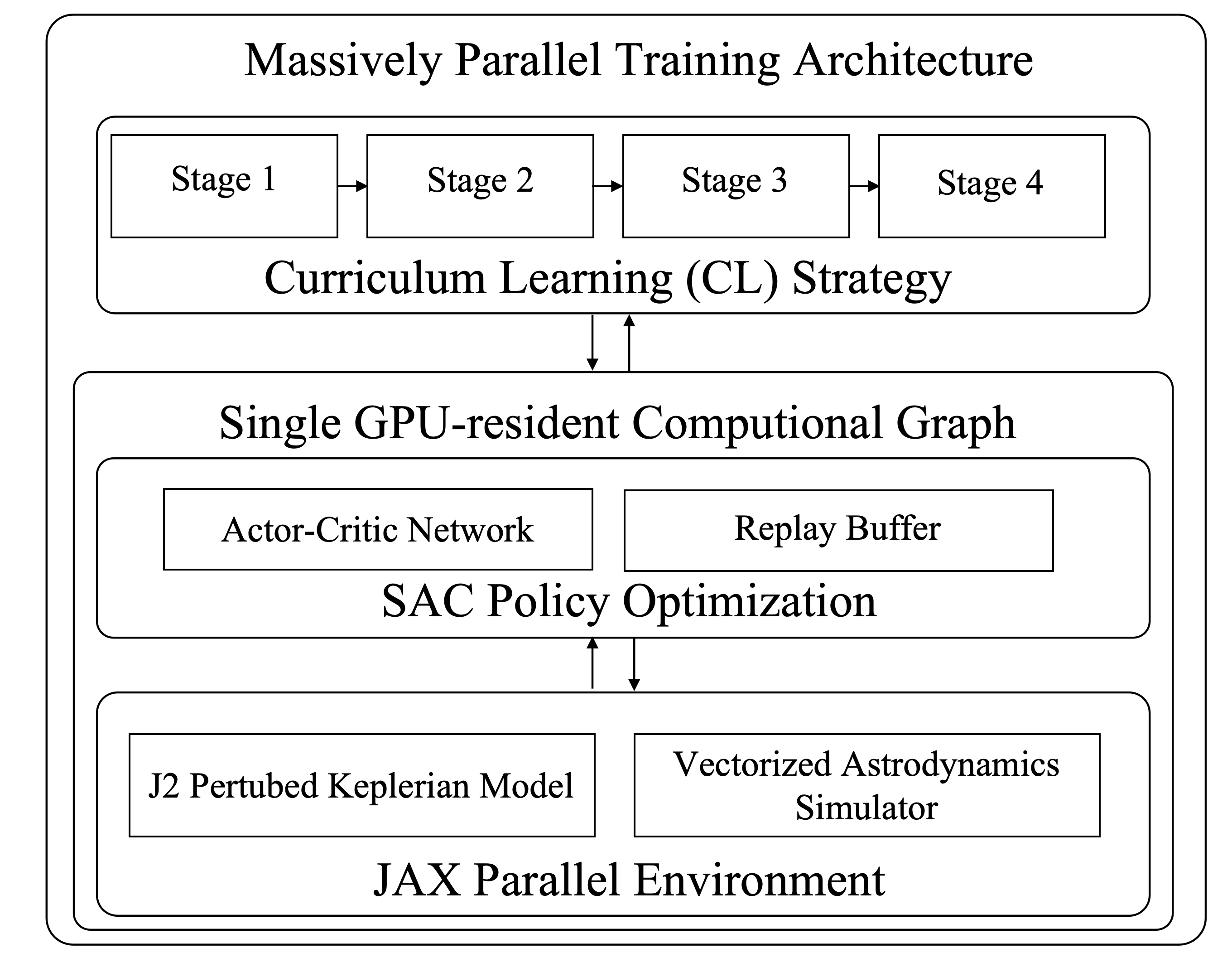}
    \caption{Schematic of the Massively Parallel Training Architecture. The framework unifies the JAX-based vectorized astrodynamics environment, SAC policy optimization, and CL stage progression within a single GPU-resident computational graph.}
    \label{fig:architecture}
\end{figure}

The training framework is depicted in Fig.\ref{fig:architecture}. The framework is composed of a massively parallel GPU-resident orbital simulator implemented via JAX, the SAC algorithm, and the CL strategy. The framework provides a practical learning approach for the long-term, multi-revolution rendezvous problem. The formulations and technical implementations are presented in the following subsections.

\paragraph{Parallel environment via JAX} To break the throughput bottleneck of traditional CPU simulations, we construct a fully tensorized, GPU-resident orbital propagator via JAX. This zero-copy architecture avoids repeated cross-device communication, generating millions of {$J_2$-perturbed state transitions} directly in GPU memory to provide the high data throughput required for long-horizon RL.
    
\paragraph{SAC algorithm} Multi-revolution rendezvous under $J_2$ perturbations presents a highly non-convex optimization landscape with numerous local optima. The off-policy nature of the SAC algorithm improves sample efficiency in conjunction with the parallel environments, while the maximum-entropy objective encourages stochastic exploration, which helps prevent the policy from prematurely collapsing into suboptimal trajectories.
    
\paragraph{Curriculum learning} The severe reward sparsity of multi-revolution rendezvous makes it nearly impossible for a randomly initialized agent to reach the target. CL strategy decomposes the task into four learning stages, providing the agent with dense and achievable goals in each stage in order to assist the agent to learn the feasible policy more effectively.

\subsection{Massively Parallel RL Framework Based on JAX}
\label{3-subsec1}
The generation of state transition data heavily relies on the high-fidelity numerical integration of the governing differential equations (e.g., Eq. \ref{eq:gve_matrix}). When simulating long-horizon rendezvous tasks that require thousands of integration steps per episode, CPU-based sequential ordinary differential equation solvers become a severe computational bottleneck. Furthermore, employing standard multi-processing technologies (such as MPI) to parallelize the environment introduces significant communication latency across the PCIe bus when transferring trajectory data from the CPU to the neural networks residing on the GPU. Consequently, traditional reinforcement learning environments (e.g., standard OpenAI Gym implementations) are often constrained by CPU computational capacity and cross-device communication overhead \cite{makoviychuk2021isaac}.

Although GPU-accelerated physics engines have emerged in recent years to alleviate this bottleneck (e.g., Isaac Gym \cite{makoviychuk2021isaac} and Brax \cite{freeman2021brax}), the vast majority of these frameworks are tailored for rigid-body robotics, articulated joint motions, and contact dynamics. They inherently lack the high-order numerical integrators and perturbation models (such as the $J_2$ gravitational effect and Modified Equinoctial Elements formulations) strictly required for high-fidelity astrodynamics \cite{vallado2001fundamentals,makoviychuk2021isaac}. Therefore, directly applying these existing robotics simulation environments to orbital mechanics yields unacceptable integration truncation errors during long-duration flight simulations.

To address this data-throughput limitation and enable large-scale parallel computation for astrodynamics, we construct a fully tensorized, domain-specific, hardware-accelerated parallel environment architecture utilizing JAX.

JAX is a numerical computing framework designed for high-performance machine learning and scientific computing, enabling composable functional transformations of native Python and NumPy programs. Unlike conventional deep learning libraries that rely on dynamic computational graphs, JAX utilizes the Accelerated Linear Algebra (XLA) compiler as its core execution backend. Through Just-In-Time (JIT) compilation, it translates tensor operations into low-level computational instructions highly optimized for GPU architectures \cite{jax2018github}.

Furthermore, JAX provides an auto-vectorization operator (\texttt{vmap}) capable of lifting serial physics computation functions---originally designed for single-satellite state propagation---into a batched format. This enables massive parallel computing without necessitating modifications to the underlying physical model's implementation.  Rooted in pure function and stateless programming paradigms, this computational model is well suited to the high-fidelity numerical integration of orbital dynamics. Concurrently, it mitigates the performance overhead typically incurred by cross-device communication during reinforcement learning training.

{The nominal training model is selected to balance physical representation and simulation throughput. The $J_2$ term captures the dominant contribution of the Earth's oblateness to the gravitational perturbation~\cite{vallado2001fundamentals}. Its analytical MEE formulation uses fixed-shape elementwise operations that map directly to JAX vectorization and XLA compilation, without external atmospheric or ephemeris queries in the rollout loop. This makes the model particularly convenient for fully GPU-resident training, retaining the leading oblateness effects while supporting high-throughput generation of training transitions. This choice does not imply that additional perturbations cannot be implemented on a GPU. Their effect on the learned guidance is assessed separately through closed-loop evaluation of the frozen policy under the high-fidelity dynamical model.}

\begin{figure}[t]%
\centering%

\includegraphics[width=0.7\linewidth]{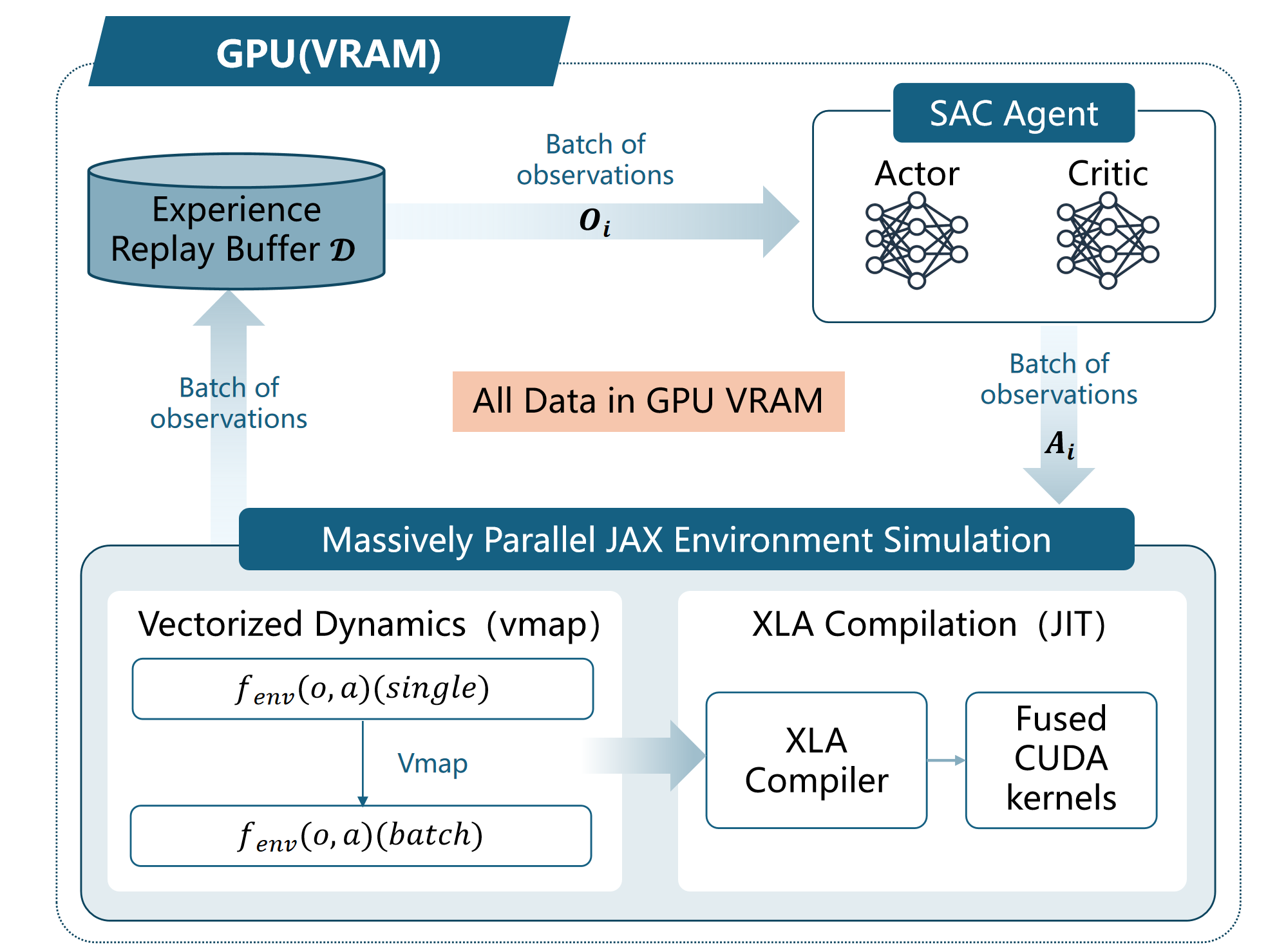}
\caption{Schematic of the Massively Parallel Training Architecture}\label{fig:parallel}
\end{figure}

In this study, the MEE dynamic model, the $J_2$ perturbation model, and the Runge–Kutta numerical integrator are all implemented as pure, stateless functions. Based on this design, a massively parallel dynamic simulation environment can be constructed within a unified tensor computation framework.The proposed JAX-based parallel training architecture is illustrated in Fig.\ref{fig:parallel}. Specifically, the system primarily adopts the following two computational implementations.

\paragraph{Vectorized dynamics and XLA compilation}
Let the discrete-time state transition function of the environment be defined as $f_{\text{env}}(\boldsymbol{o}_i, \boldsymbol{a}_i)$. Instead of relying on multi-threading concurrency across $N_E$ independent environment instances, we utilize the JAX \texttt{vmap} (vectorizing map) transformation. This operation automatically pushes the loop over the environment batch dimension down into the primitive tensor operations:
\begin{equation}
    \boldsymbol{O}_{i+1} = \text{vmap}(f_{\text{env}})(\boldsymbol{O}_i, \boldsymbol{A}_i)
\end{equation}
where $\boldsymbol{O}_i \in \mathbb{R}^{N_E \times 14}$ and $\boldsymbol{A}_i \in \mathbb{R}^{N_E \times 3}$ denote the batched observation and action tensors across $N_E$ parallel satellites, respectively. 

To maximize GPU utilization, a fixed-step fifth-order Runge--Kutta (RK5) integrator is adopted for orbital propagation. The integration step size is identical for all environments. Between two consecutive decision steps, the orbital dynamics are propagated using multiple fixed RK5 integration substeps while the applied thrust command remains unchanged. Therefore, the decision interval is equal to the RK5 substep size multiplied by the number of integration substeps. This fixed-step design avoids thread divergence that would arise if adaptive-step integrators were used, where different orbital conditions could require different numbers of function evaluations.

Subsequently, the vectorized state transition computation is JIT compiled into fused CUDA kernels via the XLA compiler. This ensures that the numerical integration of $N_E$ (e.g., $10^4$) independent orbital trajectories is executed strictly synchronously on the GPU streaming multiprocessors.

\paragraph{Zero-Copy on-device pipeline} 
Leveraging this tensorized architecture, the complete reinforcement learning training pipeline is fully encapsulated within device memory. Both the state tensors of the parallel environments and the experience replay buffer persist in GPU video random-access memory (VRAM). The entire process, including SAC action evaluation and numerical integration of the orbital dynamics, requires minimal host-to-device (i.e., CPU-to-GPU) data transfers.  This zero-copy mechanism substantially reduces wall-clock simulation time compared with traditional architectures, thereby improving the tractability of multi-revolution rendezvous exploration.

Ultimately, these two computational implementations provide practical advantages for the reinforcement learning optimization process. In traditional serial or communication-bound architectures, the primary wall-clock bottleneck arises from the propagation of environmental states. In contrast, within the proposed fully tensorized architecture, the dominant wall-clock cost shifts from environment propagation to the neural network backpropagation stage. By generating large numbers of {$J_2$-perturbed orbital state transitions} per second on a single GPU, the SAC agent can explore a broad orbital state space with substantially reduced training time. Such ultra-high data throughput not only effectively mitigates the sparse reward problem associated with long-horizon credit assignment in astrodynamics, but also enables the agent to discover non-intuitive optimal control strategies for challenging scenarios—such as large-phase-angle rendezvous—through extensive trial-and-error exploration.

\subsection{SAC Algorithm}
\label{3-subsec2}

\subsubsection{Markov Decision Process Formulation}
\label{3-2-subsubsec1}
With the parallel simulation architecture, the rendezvous problem is reformulated into a Markov Decision Process (MDP) defined by the tuple $(\mathcal{S}, \mathcal{A}, \mathcal{P}, \mathcal{R}, \gamma)$, where The state $s\in \mathcal{S}$ is defined using MEEs:
\begin{equation}
    \label{eq:state}
    s=[p, f,g,h,k,L]^T
\end{equation}
and $\gamma$ is the discount factor.  

\paragraph{Observation space}
To ensure numerical stability during neural network forward propagation, the state variables are normalized in the final observation, where the semi-latus rectum $p$ is normalized by the Earth's equatorial radius $R_e$ while the other elements remain; therefore, the normalized slow variables of MEEs at step $i$ are defined as:
\begin{equation}
    \tilde{\boldsymbol{s}}_{i} = \left[\frac{p_i}{R_e}, f_i, g_i, h_i, k_i\right]^T
\end{equation}
The original observation without noise $\tilde{\boldsymbol{o}}_i$ is defined as:
\begin{equation}
    \tilde{\boldsymbol{o}}_i =  \left[
        \tilde{\boldsymbol{s}}_{i}^T, \;
        \sin L_i, \; \cos L_i, \;
        \Delta \tilde{\boldsymbol{s}}_{i}^T, \;
        \Delta L_i, \;
        t_{\text{left}, i}
   \right]^T
\end{equation}
where $L_i$ is the true longitude, {$t_{\text{left},i}$ is the remaining number of control steps divided by the maximum number of control steps}, and {$\Delta\tilde{\boldsymbol{s}}_i=\tilde{\boldsymbol{s}}_i^{\text{target}}-\tilde{\boldsymbol{s}}_i$} represents the tracking error. Decomposing the fast variable into $\sin L_i$ and $\cos L_i$ mitigates the angular singularity. The structure of observation allows the agent to perceive the dynamics model and the current state of the satellite via the absolute components $(\tilde{\boldsymbol{s}}_{i}, \sin L_i, \cos L_i)$ while simultaneously receiving the targeting errors via the relative components $(\Delta \tilde{\boldsymbol{s}}_{i}, \Delta L_i)$.

\begingroup
The relative phase is defined consistently as
$\Delta L_i=\operatorname{wrap}_{[-\pi,\pi)}(L_i^{\mathrm{target}}-L_i)$.
Navigation errors are represented by additive errors in the estimated Cartesian position and velocity. For the target and chaser spacecraft at control update $i$,
\begin{equation}
\label{eq:observation_noise}
\begin{aligned}
\hat{\mathbf r}_{j,i}&=\mathbf r_{j,i}+\boldsymbol\eta_{r,j,i},
&\boldsymbol\eta_{r,j,i}&\sim\mathcal N(\mathbf 0,\sigma_r^2\mathbf I_3),\\
\hat{\mathbf v}_{j,i}&=\mathbf v_{j,i}+\boldsymbol\eta_{v,j,i},
&\boldsymbol\eta_{v,j,i}&\sim\mathcal N(\mathbf 0,\sigma_v^2\mathbf I_3).
\end{aligned}
\end{equation}
Here $\sigma_r$ and $\sigma_v$ are per-axis standard deviations in metres and metres per second, respectively. The measured Cartesian states are converted to MEE and normalized as above. Both the absolute and relative features of the policy observation are constructed from these measured states, while the remaining-time feature is unchanged.
\endgroup

\paragraph{Action space}
The action $\boldsymbol{u}_i \in \mathcal{A}$ corresponds to the thrust expressed in the LVLH frame. The original output of the actor network $\boldsymbol{\tilde{u}}_i \in \mathbb{R}^3$ is mapped to $[-1, 1]^3$ via a hyperbolic tangent function:
\begin{equation}
    \boldsymbol{u}_i = \tanh(\boldsymbol{\tilde{u}}_i)
\end{equation}
where $\boldsymbol{u}_i = [u_r, u_t, u_n]^T$ acts as the normalized thrust ratio for each RTN component, with each component bounded in $[-1,1]$. The final thrust applied to the orbital dynamics is then recovered via $\boldsymbol{T}_i = T_{\max} \cdot \boldsymbol{u}_i$, where $T_{\max}$ is the per-component thrust limit in the RTN frame.

\paragraph{Reward function}
The reward of step $i$ is formulated as:
\begin{equation}
    R_i = R_{\text{track}, i} + R_{\text{shape}, i} + R_{\text{fuel}, i}
\end{equation}
where the tracking penalty $R_{\text{track}, i}$, evaluates the residual errors in the MEE space:
\begin{equation}
    R_{\text{track}, i} = - \left( w_{1} \|\Delta \tilde{\boldsymbol{s}}_{i}\|_1 + w_{2} |\Delta L_i| \right)
\end{equation}
where $\|\cdot\|_1$ denotes the norm $L_1$, and $\Delta L_i \in [-\pi, \pi]$ is the wrapped angular error, the weights are fixed at $w_{1}=10$ and $w_{2}=1$.

In addition, to mitigate the sparse reward issue that occurred in low-thrust rendezvous task, a dense progress-based shaping reward is introduced to provide dense gradient signals below:
\begin{equation}
    R_{\text{shape}, i} = - w_{3} \left( \|\boldsymbol{e}_i\|_2 - \|\boldsymbol{e}_{i-1}\|_2 \right)
\end{equation}
where $||\cdot||_2$ denotes the norm $L_2$, and $\boldsymbol{e}_i$ represents the residual vector at step $i$. This term grants a positive incentive for any incremental reduction in the distance towards the target. The shaping weight is empirically fixed at $w_{3}=10$ to ensure that the magnitude of the dense reward matches the tracking penalty.

Finally, to enforce propellant efficiency, a fuel consumption penalty proportional to the control effort is applied:
\begin{equation}
    R_{\text{fuel}, i} = - \lambda \|\boldsymbol{u}_i\|_2^2
\end{equation}
where $\lambda$ corresponds to the optimal control objective defined in Eq~\eqref{eq:cost_function} and Eq~\eqref{eq:ocp_summary} and is set as 1.

\paragraph{State transition}
The state transition from step $i$ to $i+1$ is defined by the GVE formulated in Eq~\eqref{eq:j2_pert}:
\begin{equation}
    \boldsymbol{s}_{i+1} = \boldsymbol{s}_i + \int_{t_i}^{t_i + \Delta t_{\text{dec}}} \left[ \boldsymbol{B}(\boldsymbol{s}(\tau)) \cdot \left( \frac{T_{\max}}{m} \boldsymbol{u}_i + \boldsymbol{a}_{J_2}(\boldsymbol{s}(\tau)) \right) + \boldsymbol{P}(\boldsymbol{s}(\tau)) \right] d\tau
\end{equation}
where $\Delta t_{\text{dec}} = N_{\text{sub}}\Delta t$ denotes the decision interval, $\Delta t$ is the fixed RK5 integration substep, and $\boldsymbol{u}_i$ is held constant over the interval $[t_i, t_i+\Delta t_{\text{dec}}]$.

\paragraph{Success criterion}
The episode ends successfully when both the relative position and velocity between the chaser ($\boldsymbol{x}_{c}$) and the target ($\boldsymbol{x}_{d}$) meet the predefined precision thresholds:
\begin{equation}
    \begin{cases}
        \|\boldsymbol{r}_c - \boldsymbol{r}_d\|_2 \le \epsilon_r \\
        \|\boldsymbol{v}_c - \boldsymbol{v}_d\|_2 \le \epsilon_v
    \end{cases}
\end{equation}
where $\epsilon_r$ and $\epsilon_v$ are the position and velocity tolerances, respectively.

\subsubsection{Soft Actor-Critic Algorithm}
\label{3-2-subsubsec2}

The SAC algorithm is selected to solve the MDP defined in Section~\ref{3-overview} which is described in Algorithm~\ref{alg:sac}. SAC is a deep reinforcement learning algorithm based on the maximum entropy framework\cite{haarnoja2018soft, haarnoja2019softactorcriticalgorithmsapplications}, coupling the replay of off-policy experience with a maximum entropy objective, which prevents the agent from collapsing into suboptimal local optima.

\begin{algorithm}[htbp]
\caption{Soft Actor-Critic Algorithm}
\label{alg:sac}
\begin{algorithmic} %
\STATE \textbf{Initialize} soft Q-network parameters $\theta_1, \theta_2$ and target parameters $\bar{\theta}_1 \leftarrow \theta_1$, $\bar{\theta}_2 \leftarrow \theta_2$
\STATE \textbf{Initialize} policy network parameters $\phi$
\STATE \textbf{Initialize} replay buffer $\mathcal{D}$

\FOR{each iteration}
    \FOR{each environment step}
        \STATE Sample continuous action $\boldsymbol{a}_t \sim \pi_\phi(\boldsymbol{a}_t | \boldsymbol{s}_t)$
        \STATE Execute action $\boldsymbol{a}_t$, observe next state $\boldsymbol{s}_{t+1} \sim p(\boldsymbol{s}_{t+1} | \boldsymbol{s}_t, \boldsymbol{a}_t)$
        \STATE Compute reward $r_t = r(\boldsymbol{s}_t, \boldsymbol{a}_t)$
        \STATE Store transition in replay buffer: $\mathcal{D} \leftarrow \mathcal{D} \cup \{(\boldsymbol{s}_t, \boldsymbol{a}_t, r_t, \boldsymbol{s}_{t+1})\}$
    \ENDFOR
    
    \FOR{each gradient step}
        \STATE Sample a mini-batch of transitions from $\mathcal{D}$
        \STATE Update soft Q-networks for $j \in \{1, 2\}$: $\theta_j \leftarrow \theta_j - \lambda_Q \hat{\nabla}_{\theta_j} J_Q(\theta_j)$
        \STATE Update policy network: $\phi \leftarrow \phi - \lambda_\pi \hat{\nabla}_\phi J_\pi(\phi)$
        \STATE Soft update target Q-networks for $j \in \{1, 2\}$: $\bar{\theta}_j \leftarrow \tau \theta_j + (1 - \tau)\bar{\theta}_j$
    \ENDFOR
\ENDFOR
\end{algorithmic}
\end{algorithm}

\paragraph{Maximum entropy objective}
For the policy $\pi$ from the actor network, the SAC algorithm augments this objective with an entropy term $\mathcal{H}$, designed to keep the policy stochastic:
\begin{equation}
    J(\pi) = \mathbb{E}_{\tau \sim \pi} \left[ \sum_{t=0}^{T} \gamma^t \left( R(s_t, a_t) + \alpha \mathcal{H}(\pi(\cdot | s_t)) \right) \right]
\end{equation}
where $\alpha$ is the temperature parameter that determines the relative importance of the entropy term to the reward. In this paper, $\alpha$ is automatically tuned during training to maintain a target entropy.

\paragraph{Critic update}
SAC makes use of two soft Q-functions ($Q_{\theta_1}, Q_{\theta_2}$) to mitigate positive bias in the policy improvement step. The parameters $\theta$ are optimized to minimize the soft Bellman residual:
\begin{equation}
    J_Q(\theta) = \mathbb{E}_{(s_t, a_t) \sim \mathcal{D}} \left[ \frac{1}{2} \left( Q_\theta(s_t, a_t) - \hat{Q}(s_t, a_t) \right)^2 \right]
\end{equation}
where $\mathcal{D}$ is the replay buffer. The target $\hat{Q}$ is computed using the target Q-networks $(Q_{\bar{\theta}_1}, Q_{\bar{\theta}_2})$ and the current policy:
\begin{equation}
    \hat{Q}(s_t, a_t) = r_t + \gamma \left( \min_{j=1,2} Q_{\bar{\theta}_j}(s_{t+1}, a_{t+1}) - \alpha \log \pi_\phi(a_{t+1}|s_{t+1}) \right)
\end{equation}

\paragraph{Actor update}
The policy network $\pi_\phi$ parameterized by $\phi$ models a Gaussian distribution by predicting its mean and covariance. To enable gradient backpropagation through the stochastic sampling process, the action is defined as $\boldsymbol{a}_t = f_\phi(\boldsymbol{\epsilon}_t; \boldsymbol{s}_t)$, where $\boldsymbol{\epsilon}_t \sim \mathcal{N}(\mathbf{0}, \mathbf{I})$ is sampled from a standard multivariate normal distribution. The policy parameters are updated by minimizing the entropy-regularized objective function:
\begin{equation}
    J_\pi(\phi) = \mathbb{E}_{\boldsymbol{s}_t \sim \mathcal{D},\, \boldsymbol{\epsilon}_t \sim \mathcal{N}(\mathbf{0}, \mathbf{I})} \left[ \alpha \log \pi_\phi \big( f_\phi(\boldsymbol{\epsilon}_t; \boldsymbol{s}_t) \mid \boldsymbol{s}_t \big) - \min_{j=1,2} Q_{\theta_j} \big( \boldsymbol{s}_t, f_\phi(\boldsymbol{\epsilon}_t; \boldsymbol{s}_t) \big) \right] \label{eq:policy_loss}
\end{equation}
This formulation naturally balances exploitation and exploration, driving the policy toward state-action spaces with higher expected values while preserving sufficient stochasticity to prevent premature convergence and ensure robust guidance performance.

\subsection{Curriculum Learning Strategy}

\label{3-4}

The four stages are detailed in table \ref{tab:curriculum_levels}. After every $K$ training episode, the agent is evaluated on 100 randomly generated test cases within the current difficulty range. The curriculum advances to the next level only if the success rate of current level exceeds $90\%$.

\begin{table}[htbp]
\centering
\caption{Curriculum Learning Stages and Orbital Element Variation Ranges}
\label{tab:curriculum_levels}

\begin{tabular}{@{}lcccc@{}}
\toprule
\textbf{Parameter} &
\textbf{Stage 1} &
\textbf{Stage 2} &
\textbf{Stage 3} &
\textbf{Stage 4} \\
\midrule

$\Delta a$ (km)
& $\pm 10$
& $\pm 25$
& $[-50,\,+100]$
& $[-300,\,+500]$ \\

$e$
& $[0,\,0.005]$
& $[0,\,0.01]$
& $[0,\,0.1]$
& $[0,\,0.1]$ \\

$\Delta i$ ($^\circ$)
& $\pm 0.1$
& $\pm 0.1$
& $\pm 0.2$
& $\pm 0.4$ \\

$\Delta \Omega$ ($^\circ$)
& $\pm 1$
& $\pm 1$
& $\pm 1$
& $\pm 1$ \\

$\Delta \omega$ ($^\circ$)
& $0$
& $0$
& $0$
& $\pm 180$ \\

$\Delta \nu$ ($^\circ$)
& $[0,180]$
& $[0,180]$
& $[-180,180]$
& $[-180,180]$ \\

\bottomrule
\end{tabular}

\end{table}

Besides, a re-specialization strategy is proposed each time the curriculum level advances. The parameters of the neural networks are transferred to the next stage, while the replay buffer $\mathcal{D}$ and the temperature parameter $\alpha$ are reset. Replay buffer resetting removes the experiences collected from previous stages which correspond to easier tasks, preventing the agent from overfitting to simple tasks and forcing it to collect new experiences relevant to the harder task distribution. Resetting $\alpha$ reactivates the exploration mechanism, which is crucial when the agent encounters the expanded state space of the new curriculum level. The strategy balances the retention of control policies with the plasticity required to adapt to increasingly complex rendezvous tasks.

To progressively improve the generalization capability of the policy, the curriculum stages are associated with increasingly broader orbital-element distributions. The initial orbital altitude is randomly sampled within the low-Earth-orbit regime, covering approximately 450--2000 km in the most challenging stage. Furthermore, physically motivated constraints are imposed during task generation. For a given semi-major axis $a$, the eccentricity is sampled such that the resulting perigee altitude always satisfies $r_p=a(1-e)\geq R_E+400\,\mathrm{km}$. This constraint automatically excludes physically infeasible orbits and ensures that all generated rendezvous scenarios remain realistic while exposing the agent to increasingly diverse orbital configurations.

\section{Simulation and Results}
\label{sec:results}

\subsection{Overview and Experimental Setup}
\label{subsec:setup}

This section validates the proposed method through three categories of experiments.
\paragraph{Optimality and robustness} {The proposed method is compared with the direct and hybrid methods under the nominal $J_2$ model in terms of time of flight, fuel consumption, and terminal-state accuracy. Robustness is assessed separately through Monte Carlo evaluations under the high-fidelity model, including eclipse-induced thrust interruptions and navigation errors.}
\paragraph{Ablation studies} Ablation studies are conducted by isolating the curriculum learning strategy. The role of curriculum learning is rigorously demonstrated by contrasting training convergence curves, empirical success rates, and trajectory optimality.
\paragraph{Computational efficiency} The raw simulation throughput of the JAX-based GPU architecture is directly compared against multi-core CPU setups. In addition, the execution time of the proposed method is evaluated against traditional numerical solvers to confirm its real-time operational capacity.
\\

\paragraph{{Common parameters and comparison methods}}
The parameters of the chaser satellite and the geographical constants are listed in Table \ref{tab:physics_params}. The initial COEs of the chaser are uniformly sampled from predefined bounded intervals detailed in Table \ref{tab:mission_params} at the beginning of each episode. Subsequently, the initial COEs of the target are generated relative to the initial state of the chaser to represent practical orbital transfer constraints. The hyperparameters of the parallel training environment and for training process are detailed in Table \ref{tab:rl_params}.

\begin{table}[htbp]
\centering
\caption{Chaser parameters and geographical constants.}
\label{tab:physics_params}
\begin{tabular}{@{}llc@{}}
\toprule
\textbf{Parameter} & \textbf{Symbol} & \textbf{Value} \\ \midrule
Earth gravitational parameter & $\mu$ & $3.986 \times 10^{14}$ $\text{m}^3/\text{s}^2$ \\
Earth equatorial radius & $R_e$ & $6378.137$ km \\
$J_2$ perturbation coefficient & $J_2$ & $1.0826 \times 10^{-3}$ \\
Chaser initial mass & $m_0$ & $100$ kg \\
Maximum thrust per RTN component & $T_{\text{max}}$ & $0.05$ N \\
Specific impulse & $I_{\text{sp}}$ & $3000$ s \\
RK5 integration step & $\Delta t$ & $10$ s \\
Decision substeps & $N_{\text{sub}}$ & $80$ \\
Decision interval & $\Delta t_{\text{dec}}$ & $800$ s \\
\bottomrule
\end{tabular}
\end{table}

\begin{table*}[htbp]
\centering

\caption{Sampling bounds for the initial chaser state and the relative target state.}
\label{tab:mission_params}

\begin{tabular}{@{}llcc@{}}
\toprule
\textbf{Orbital Element} &
\textbf{Symbol} &
\textbf{Chaser Initial Bounds (Uniform)} &
\textbf{Target Relative Bounds} \\
\midrule

Semi-major axis
& $a$
& $[R_e+450,\;R_e+2000]$ km
& $[-300,+500]$ km \\

Eccentricity
& $e$
& $[0,\;0.1]^{\dagger}$
& $[0,\;0.1]^{\dagger}$ \\

Inclination
& $i$
& $[40^\circ,\;80^\circ]$
& $\pm0.4^\circ$ \\

Right ascension of ascending node
& $\Omega$
& $[0^\circ,\;360^\circ]$
& $\pm1^\circ$ \\

Argument of perigee
& $\omega$
& $[0^\circ,\;360^\circ]$
& $\pm180^\circ$ \\

True anomaly
& $\nu$
& $[0^\circ,\;360^\circ]$
& $\pm180^\circ$ \\

\bottomrule
\end{tabular}

\vspace{2mm}

\parbox{0.95\textwidth}{\footnotesize
$^{\dagger}$ For the target orbit, the semi-major axis is first generated subject to a minimum orbital altitude constraint of 400 km. The eccentricity is then sampled such that the corresponding perigee radius satisfies
$r_p=a(1-e)\geq R_e+400\,\mathrm{km}$.
Consequently, all generated target orbits satisfy both the minimum-altitude and minimum-perigee-altitude requirements, ensuring physically feasible rendezvous scenarios.
}

\end{table*}

\begin{table}[htbp]
\centering
\caption{Hyperparameters of the parallel training environment and for training process.}
\label{tab:rl_params}
\begin{tabular}{@{}llc@{}}
\toprule
\textbf{Hyperparameter} & \textbf{Symbol} & \textbf{Value} \\ \midrule
Number of parallel environments & $N_E$ & $320$ \\
Actor/Critic network architecture & - & $[512, 512, 512]$ \\
Activation function & - & ReLU \\
Learning rate (Actor) & $\eta_\pi$ & $3 \times 10^{-4}$ \\
Learning rate (Critic) & $\eta_Q$ & $1 \times 10^{-3}$ \\
Discount factor & $\gamma$ & $0.995$ \\
Target smoothing coefficient & $\tau$ & $0.005$ \\
Entropy temperature & $\alpha$ & Auto-tuned \\
Replay buffer capacity & $|\mathcal{D}|$ & $5 \times 10^6$ transitions \\
Batch size & $B$ & $4096$ \\
\bottomrule
\end{tabular}
\end{table}

{The following strategies are used in the nominal-model optimality comparisons and curriculum-learning ablation:}
\begin{itemize}
    \item \textbf{Proposed Method}: The agent trained using the method proposed in this paper.
    \item \textbf{SAC-NoCL}: A standard SAC agent trained directly in Stage 4 without curriculum guidance, which is utilized to demonstrate the necessity of CL in Sec~\ref{subsec:ablation}.
    \item \textbf{Direct Method}: A Gaussian pseudo-spectral optimization software (GPOPS-II)~\cite{gpops2}. To ensure NLP convergence, the solver is initialized using RQ-Law~\cite{naray2023equinoctical}.
    \item \textbf{Hybrid Method}: A sequential approach where the continuous trajectory generated by the Proposed Method serves as the initial guess for GPOPS-II, demonstrating the practical value of the RL policy as a robust warm-start generator.
\end{itemize}

\begingroup
\paragraph{Settings for robustness evaluation}\label{par:robustness_settings}
The frozen policy is evaluated in closed loop under the high-fidelity model in Section~\ref{subsec:robustness}, without retraining. Training and the other experiments use the nominal $J_2$ model. Both spacecraft are propagated under the high-fidelity model, with control applied only to the chaser.

For the target and chaser spacecraft, the perturbation acceleration comprises
\begin{equation}
\mathbf a_{\mathrm{pert},j}=\mathbf a_{J_2,j}+\mathbf a_{D,j}
+\mathbf a_{\mathrm{Sun},j}+\mathbf a_{\mathrm{Moon},j}+\mathbf a_{\mathrm{SRP},j}.
\end{equation}
In inertial coordinates, the additional accelerations are
\begin{align}
\mathbf a_{D,j}&=-\frac{\rho_j C_{D,j}A_{D,j}}{2m_j}
 \|\mathbf v_{\mathrm{rel},j}\|\mathbf v_{\mathrm{rel},j},\\
\mathbf a_{b,j}&=\mu_b\left(\frac{\mathbf R_b-\mathbf r_j}{\|\mathbf R_b-\mathbf r_j\|^3}
 -\frac{\mathbf R_b}{\|\mathbf R_b\|^3}\right),\quad b\in\{\mathrm{Sun},\mathrm{Moon}\},\\
\mathbf a_{\mathrm{SRP},j}&=\chi_j P_0\frac{C_{R,j}A_{\mathrm{SRP},j}}{m_j}
 \left(\frac{\mathrm{AU}}{\|\mathbf r_j-\mathbf R_{\mathrm{Sun}}\|}\right)^2
 \frac{\mathbf r_j-\mathbf R_{\mathrm{Sun}}}{\|\mathbf r_j-\mathbf R_{\mathrm{Sun}}\|}.
\end{align}
Here $\mathbf R_b$ is the geocentric position of body $b$, $\mathbf v_{\mathrm{rel},j}$ is the spacecraft velocity relative to the co-rotating atmosphere, and $\chi_j$ is zero inside the cylindrical Earth shadow and one otherwise. Density is computed using NRLMSISE-00~\cite{picone2002nrl} with a rigidly co-rotating, wind-free atmosphere; solar and lunar positions use DE440s~\cite{park2021de440}. Solar radiation pressure follows a cannonball model with constant area and reflection coefficient. Accelerations are projected into RTN for MEE propagation. Both spacecraft use the mass in Table~\ref{tab:physics_params}, which also specifies the thrust limit and integration settings; additional parameters appear in Table~\ref{tab:high_fidelity_params}.

\begin{table}[htbp]
\centering
\caption{{Additional parameters and settings for the high-fidelity closed-loop evaluation.}}
\label{tab:high_fidelity_params}
\small
\begin{tabular}{@{}p{0.35\linewidth}p{0.59\linewidth}@{}}
\toprule
Parameter & Setting \\ \midrule
Initial epoch & 2020-01-01 00:00:00 UTC \\
Atmospheric inputs & Fixed $F_{10.7}=150$, mean $F_{10.7}=150$, and daily $A_p=4$ \\
Earth orientation & Fixed DUT1 $=0$ s and polar motion $x_p=y_p=0$ \\
Drag properties & $A_{D,c}=A_{D,d}=1$ m$^2$; $C_{D,c}=C_{D,d}=2.2$ \\
Radiation properties & $A_{\mathrm{SRP},c}=A_{\mathrm{SRP},d}=1$ m$^2$; $C_{R,c}=C_{R,d}=1.3$ \\
Radiation pressure & $P_0=4.56\times10^{-6}$ N/m$^2$ at 1 AU \\
Evaluation horizon & 10000 control intervals, or $8\times10^6$ s \\
Monte Carlo conditions & 500 paired initial conditions; seed 42; final curriculum stage \\
Observation errors & $\sigma_r=20$ m and $\sigma_v=0.1$ m/s per axis of each spacecraft \\
\bottomrule
\end{tabular}
\end{table}

Atmospheric calculations use GCRS-to-ITRS transformations and WGS84 coordinates, with TAI propagation time converted to TDB for ephemeris queries. Fixed weather and Earth-orientation inputs and the inertial-$z$ approximation for $J_2$ are retained.

\paragraph{Position and velocity observation errors}
Training uses noise-free observations, whereas observation errors are introduced only during the high-fidelity robustness evaluations, with the policy kept frozen.
Observation errors follow Eq.~\eqref{eq:observation_noise}, with independent, zero-mean Gaussian errors of $\sigma_r=20$ m and $\sigma_v=0.1$ m/s per spacecraft per inertial axis. Noise is sampled once per decision interval to construct the 14-dimensional observation; propagation and success checks use true states. The six configurations cover two thrust-availability settings and three position tolerances, using the same 500 initial conditions and matched noise realizations. The tests exclude persistent bias, temporal correlation, and navigation filtering.
\endgroup

\subsection{Optimality and Robustness}
\label{subsec:performance}

{Robustness is evaluated under the high-fidelity dynamical model, including eclipse-induced thrust interruptions and navigation errors. Optimality comparisons are conducted separately under the nominal $J_2$ model.}

\subsubsection{{Robustness under the High-Fidelity Model}}
\label{subsec:robustness}

\begingroup
The frozen policy was evaluated on the same 500 initial rendezvous conditions under six high-fidelity configurations, all with the observation errors specified above. Success follows the criterion in Section~\ref{3-2-subsubsec1}. Without eclipse-induced thrust interruptions, success rates are 88.4\%, 99.4\%, and 100.0\% for the three cases listed in Table~\ref{tab:random_tests}, respectively. Earth shadowing of solar radiation pressure remains active in this configuration.
\endgroup

\begin{table}[htbp]
\centering
\caption{{Sampled rendezvous success under the high-fidelity model without eclipse-induced thrust interruptions. Noisy observations use $\sigma_r=20$ m and $\sigma_v=0.1$ m/s per axis of each spacecraft.}}
\label{tab:random_tests}
\begin{tabular}{ccc}
\toprule
$\epsilon_r$ (m) & $\epsilon_v$ (m/s) & Successes / cases (rate) \\ \midrule
100 & 10 & 442/500 (88.4\%) \\
200 & 10 & 497/500 (99.4\%) \\
300 & 10 & 500/500 (100.0\%) \\
\bottomrule
\end{tabular}
\end{table}

To further evaluate the robustness of the learned policy under practical mission constraints, an additional Monte Carlo campaign was conducted with eclipse-induced thrust interruptions.

\begin{figure}[htbp]
\centering
 \includegraphics[width=0.48\linewidth]{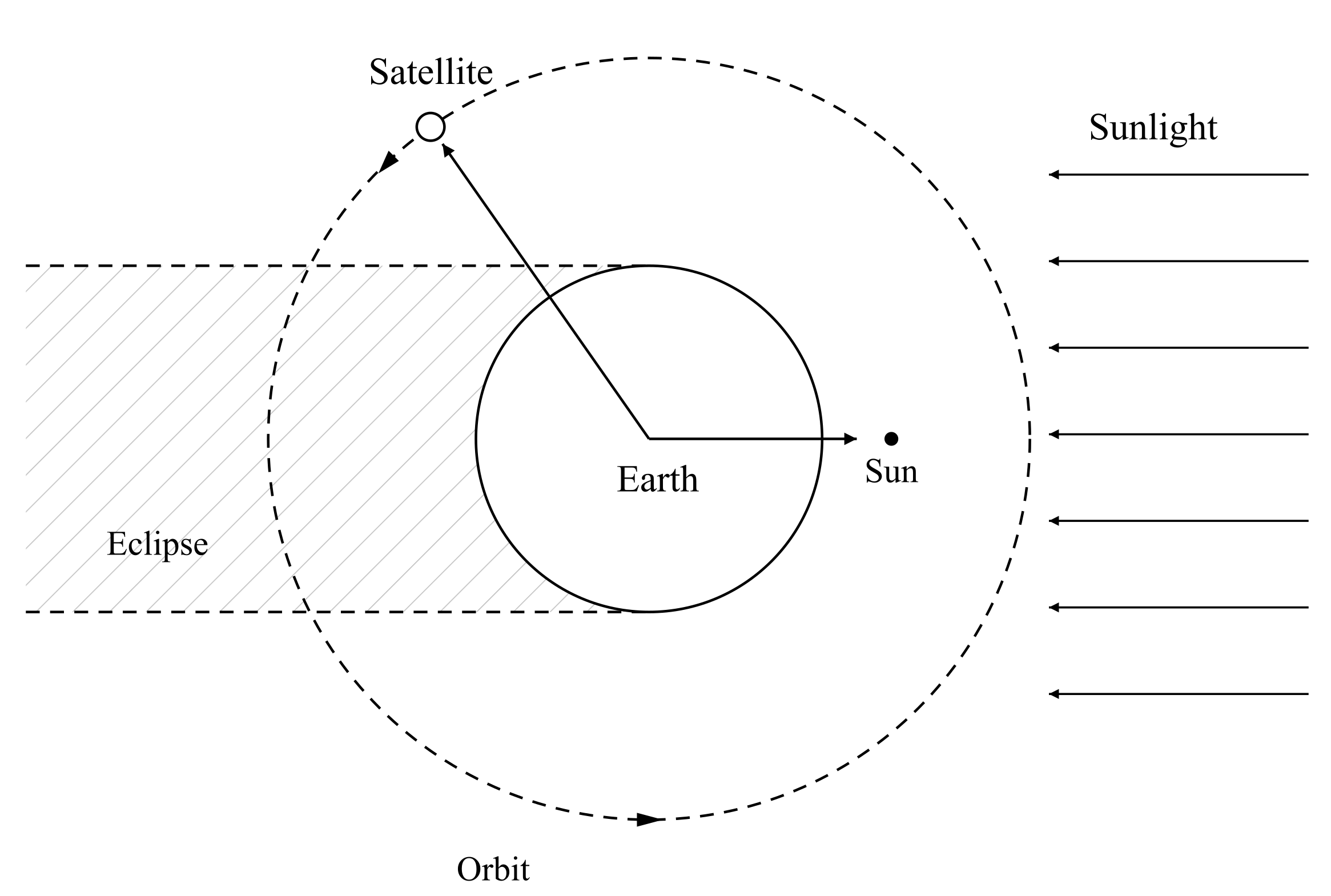}
\caption{Geometry of the cylindrical Earth-shadow model used for eclipse detection. The spacecraft is considered to be in eclipse when it is located behind the Earth with respect to the Sun and its perpendicular distance from the Sun-Earth axis is smaller than the Earth's radius.}
\label{fig:eclipse}
\end{figure}

Figure~\ref{fig:eclipse} illustrates the geometric configuration of the cylindrical Earth-shadow model adopted in this study. Let $\mathbf{r}$ denote the spacecraft position vector in the Earth-centered inertial frame and $\hat{\mathbf{s}}$ denote the unit vector pointing from the Earth toward the Sun. {The unit vector $\hat{\mathbf s}(t)=\mathbf R_{\mathrm{Sun}}(t)/\|\mathbf R_{\mathrm{Sun}}(t)\|$ is updated from DE440s at each integration stage. The cylindrical approximation is retained, without penumbra transitions.} The spacecraft is considered to be in eclipse when

\begin{equation}
\mathbf{r}^{\mathrm T}\hat{\mathbf{s}} < 0,
\end{equation}

and

\begin{equation}
\left|\mathbf{r}-
(\mathbf{r}^{\mathrm T}\hat{\mathbf{s}})\hat{\mathbf{s}}
\right| < R_e.
\end{equation}

{When eclipse-induced thrust interruptions are enabled, the applied thrust is modeled as}

\begin{equation}
\boldsymbol{T}_{\mathrm{applied}}(t)=
\begin{cases}
\boldsymbol{T}_{\mathrm{cmd}}(t), & \text{sunlight},\\
\boldsymbol{0}, & \text{eclipse}.
\end{cases}
\end{equation}
{Here $\boldsymbol{T}_{\mathrm{cmd}}(t)=T_{\max}\boldsymbol{u}_i$ is the policy-commanded thrust during decision interval $i$.}

{When the eclipse constraint is enabled, thrust is unavailable inside the shadow while gravitational and atmospheric perturbations remain active. Solar radiation pressure is suppressed in the shadow in both the constrained and unconstrained thrust configurations.}

\begingroup
With eclipse-induced thrust interruptions, success rates are 76.2\%, 94.8\%, and 98.0\% at the 100, 200, and 300 m position tolerances, respectively (Table~\ref{tab:eclipse}). The results indicate high empirical success rates at 200--300 m under the combined high-fidelity dynamics and observation errors, while the lower rates at 100 m indicate limited terminal precision.

\endgroup

\begin{table}[htbp]
\centering
\caption{{Sampled rendezvous success under the high-fidelity model with eclipse-induced thrust interruptions. Noisy observations use $\sigma_r=20$ m and $\sigma_v=0.1$ m/s per axis of each spacecraft.}}
\label{tab:eclipse}
\begin{tabular}{ccc}
\toprule
$\epsilon_r$ (m) & $\epsilon_v$ (m/s) & Successes / cases (rate) \\ \midrule
100 & 10 & 381/500 (76.2\%) \\
200 & 10 & 474/500 (94.8\%) \\
300 & 10 & 490/500 (98.0\%) \\
\bottomrule
\end{tabular}
\end{table}

\newpage
\subsubsection{{Optimality Comparison under the $J_2$ Model}}
\label{subsec:optimality}

{Two additional benchmark cases are used to compare the proposed, direct, and hybrid methods under the nominal $J_2$ model. Their initial COEs are listed in Table~\ref{tab:test_cases}. The optimization comparisons, associated trajectories, and curriculum-learning ablations retain their original nominal-model settings and are distinct from the enhanced-model Monte Carlo campaign.}

\begin{table}[htbp]
\centering
\caption{Initial COEs for the evaluated cases.}
\label{tab:test_cases}
\begin{tabular}{@{}llcccc@{}}
\toprule
\textbf{Orbital Element} & \textbf{Symbol} & \multicolumn{2}{c}{\textbf{Case 1}} & \multicolumn{2}{c}{\textbf{Case 2}} \\ \cmidrule(lr){3-4} \cmidrule(lr){5-6}
& & \textbf{Chaser} & \textbf{Target} & \textbf{Chaser} & \textbf{Target} \\ \midrule
Semi-major axis & $a$ (km) & 6898.77 & 6867.09 & 7876.03 & 8278.94 \\
Eccentricity & $e$ & $1.47\times 10^{-4}$ & $7.00\times 10^{-4}$ & 0.05 & 0.02 \\
Inclination & $i$ ($^\circ$) & 50.20 & 50.33 & 64.38 & 64.01 \\
RAAN & $\Omega$ ($^\circ$) & 28.52 & 28.47 & 67.93 & 67.64 \\
Argument of perigee & $\omega$ ($^\circ$) & 60.10 & 49.14 & 123.88 & 123.45 \\
True anomaly & $\nu$ ($^\circ$) & 145.40 & 215.75 & 96.91 & 179.19 \\ \bottomrule
\end{tabular}
\end{table}

\begin{figure}[htbp]
    \centering
    \subfloat[Proposed Method]{
        \includegraphics[width=0.47\linewidth]{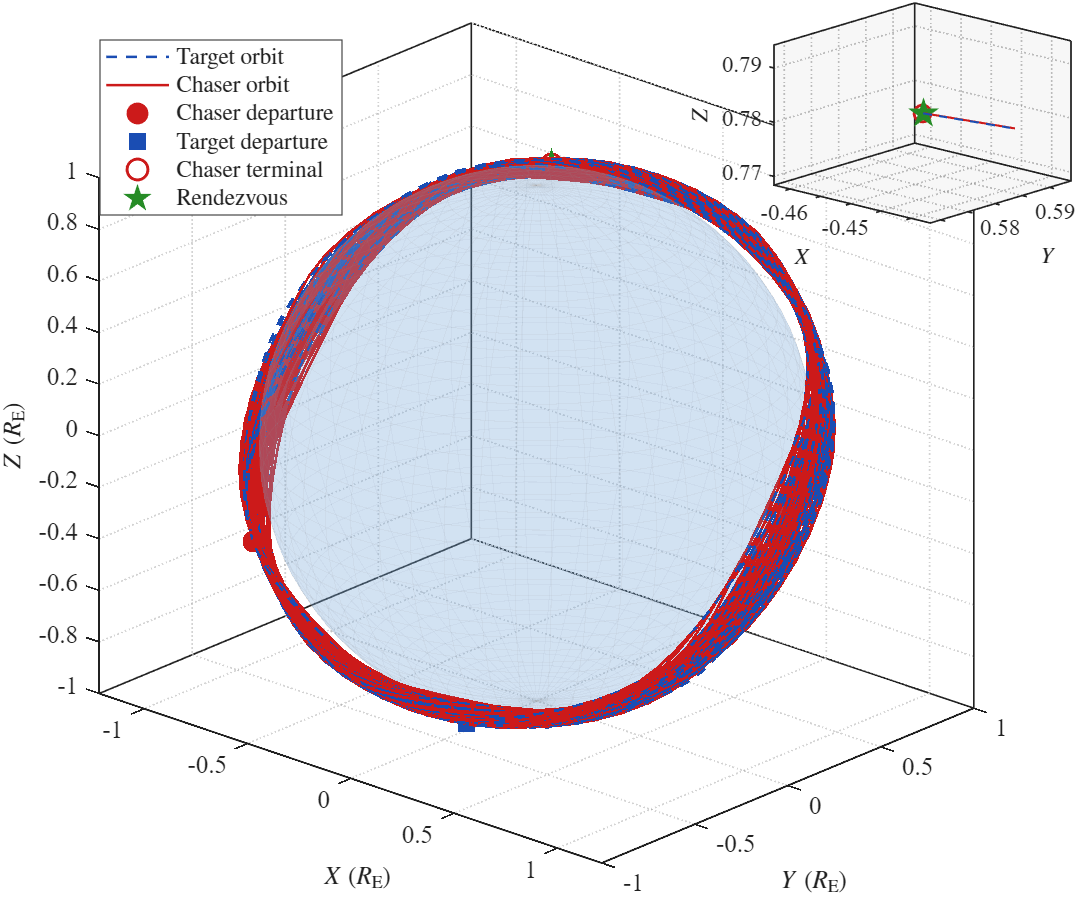}
        \label{fig:traj_proposed}
    }\hfill
    \subfloat[Hybrid (Proposed + GPOPS-II)]{
        \includegraphics[width=0.48\linewidth]{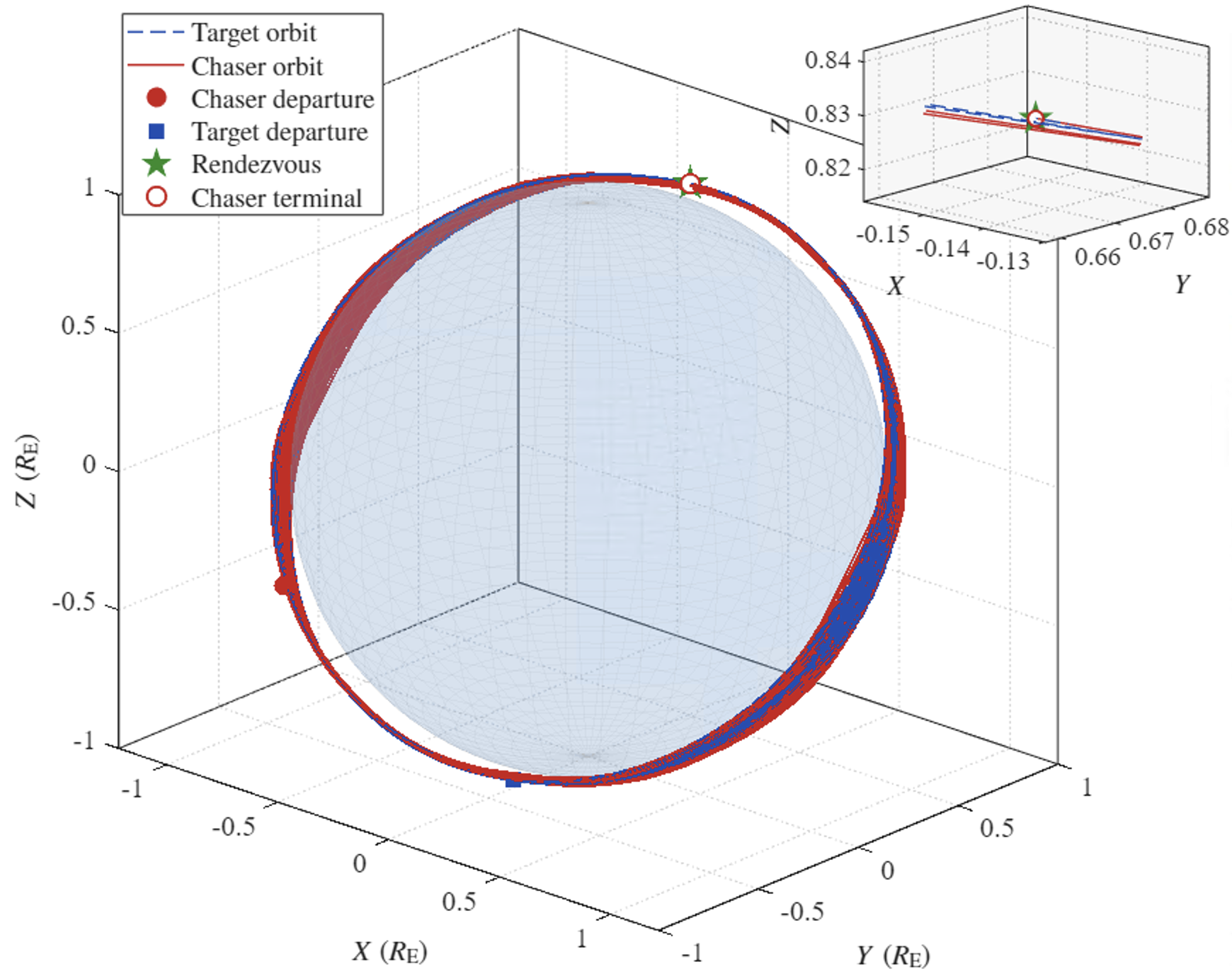}
        \label{fig:traj_hybrid}
    }
    \caption{Trajectory of proposed method and hybrid method.}
    \label{fig:trajectory_comp}
\end{figure}

\begin{figure}[htbp]
    \centering
    \subfloat[Element $p$ (km)]{
        \includegraphics[width=0.32\linewidth]{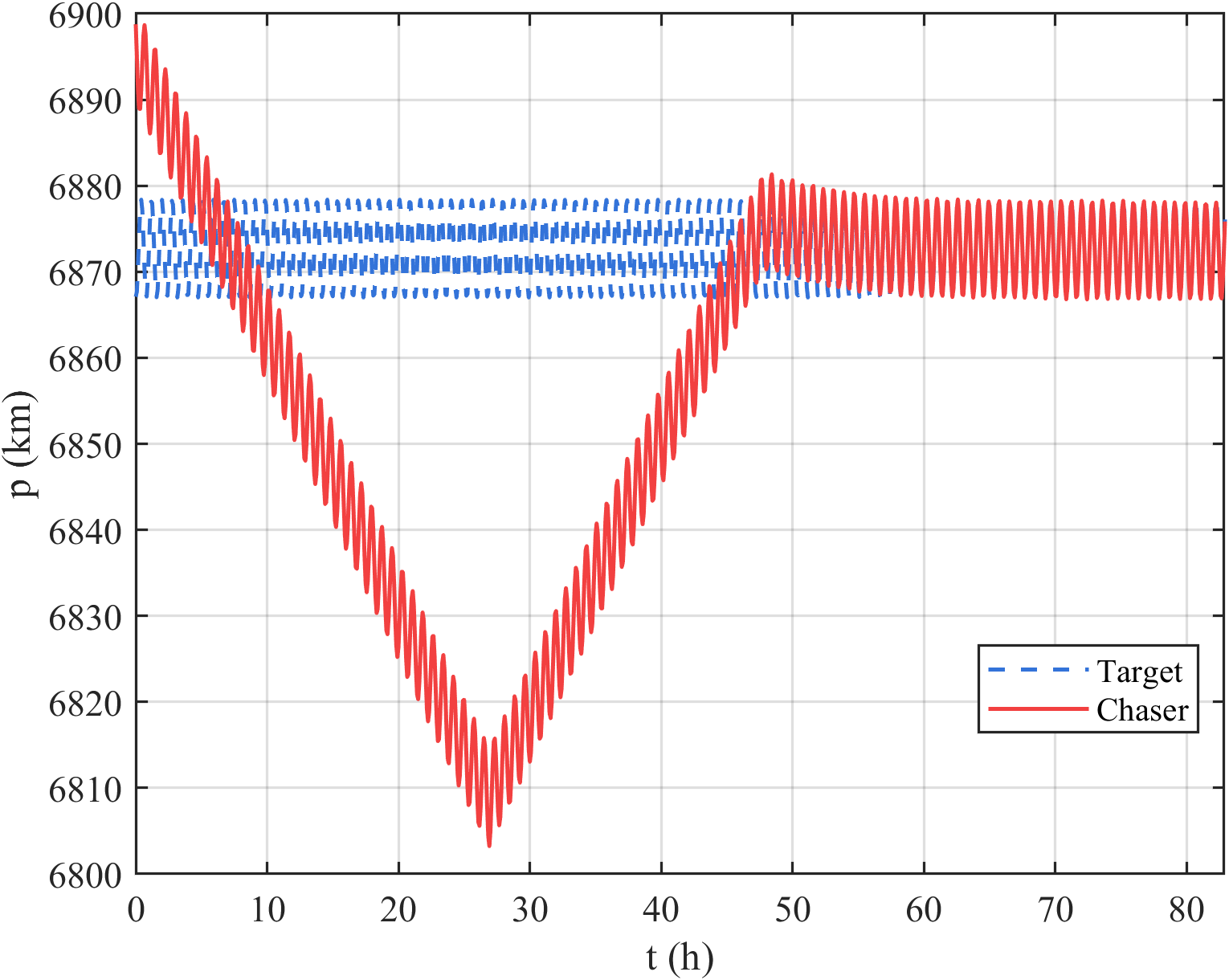}
        \label{fig:mee_p}
    }\hfill
    \subfloat[Element $f$]{
        \includegraphics[width=0.32\linewidth]{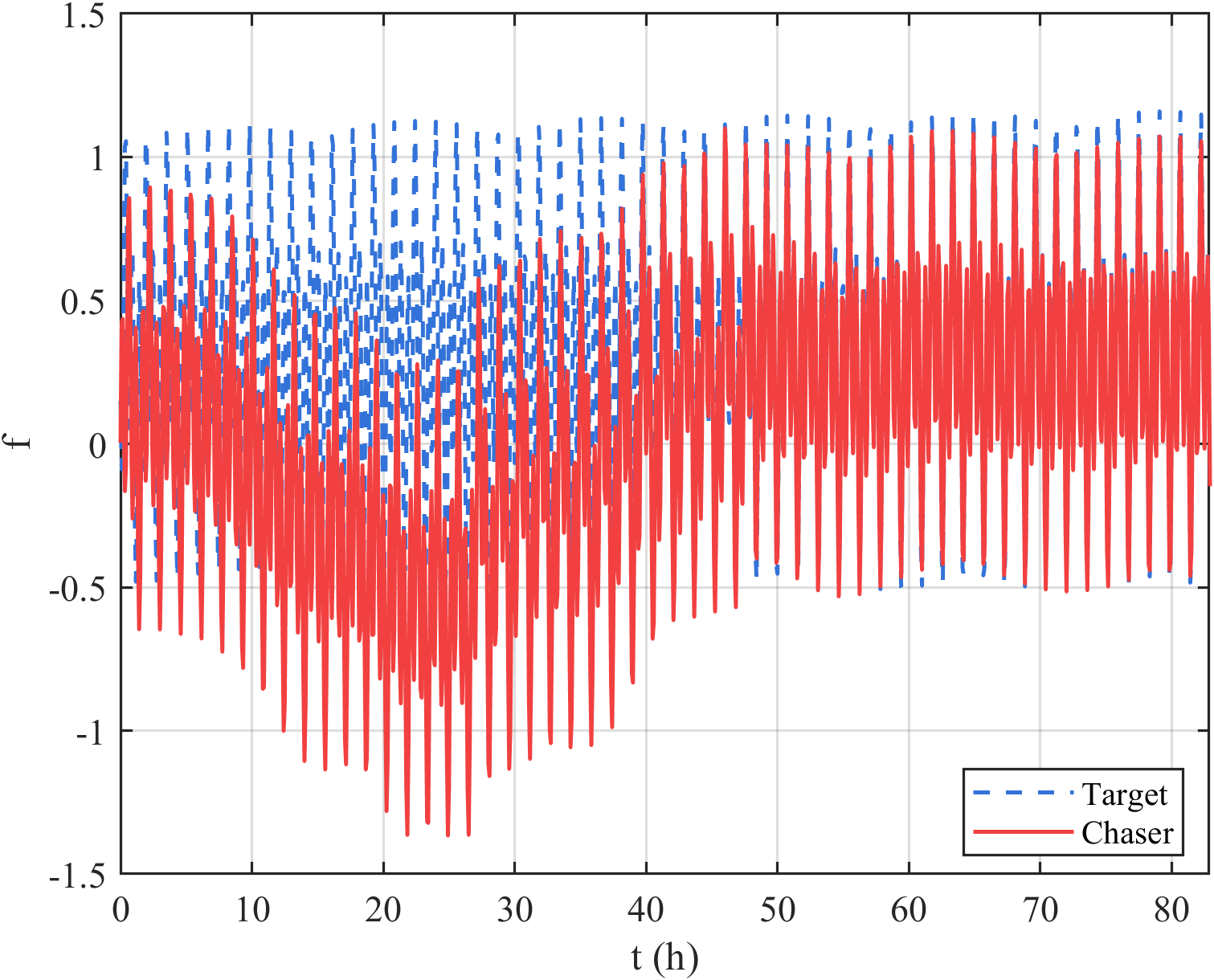}
        \label{fig:mee_f}
    }\hfill
    \subfloat[Element $g$]{
        \includegraphics[width=0.32\linewidth]{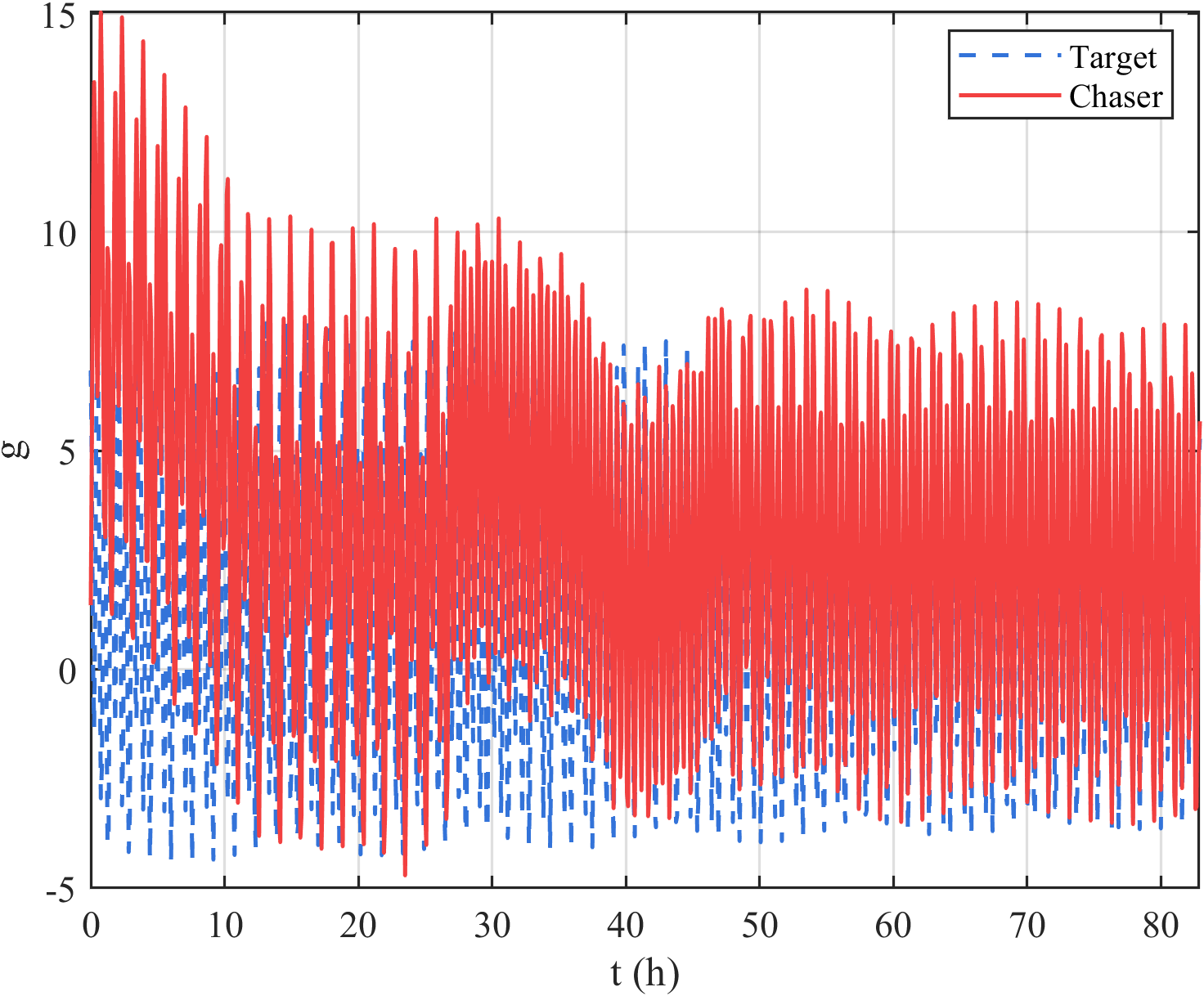}
        \label{fig:mee_g}
    }\hfill
    \subfloat[Element $h$]{
        \includegraphics[width=0.32\linewidth]{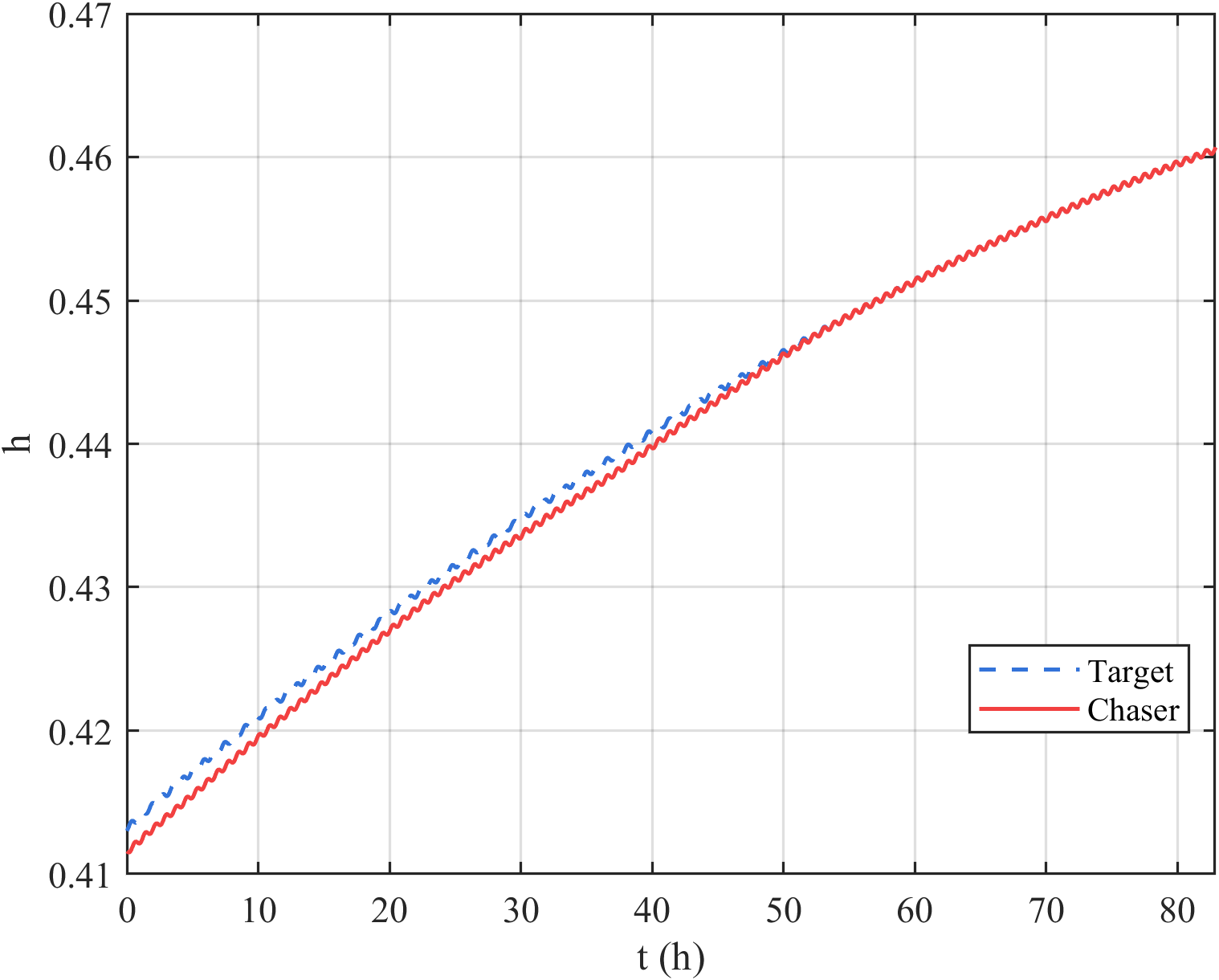}
        \label{fig:mee_h}
    }\hfill
    \subfloat[Element $k$]{
        \includegraphics[width=0.32\linewidth]{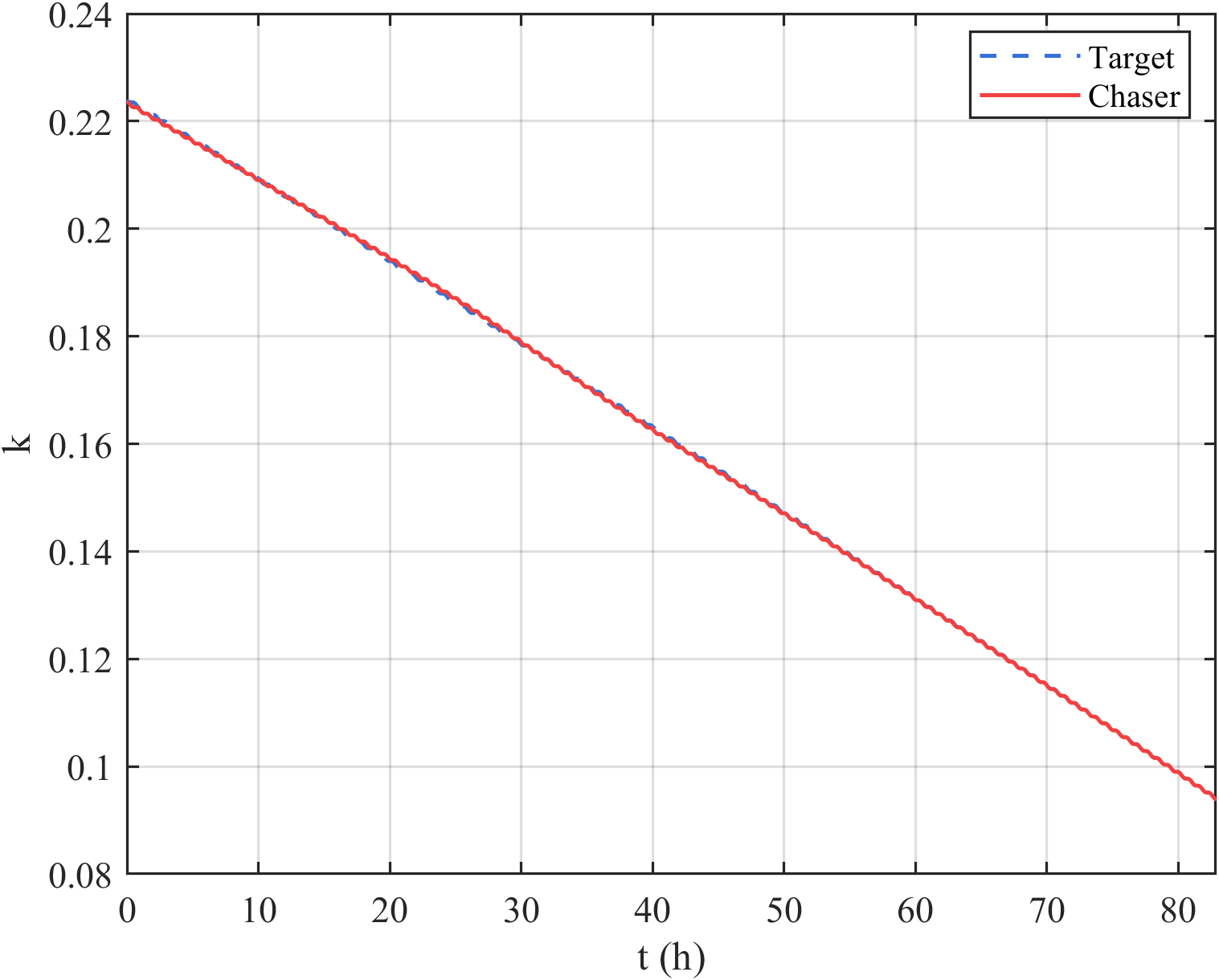}
        \label{fig:mee_k}
    }\hfill
    \subfloat[True Longitude $L$ (rad)]{
        \includegraphics[width=0.32\linewidth]{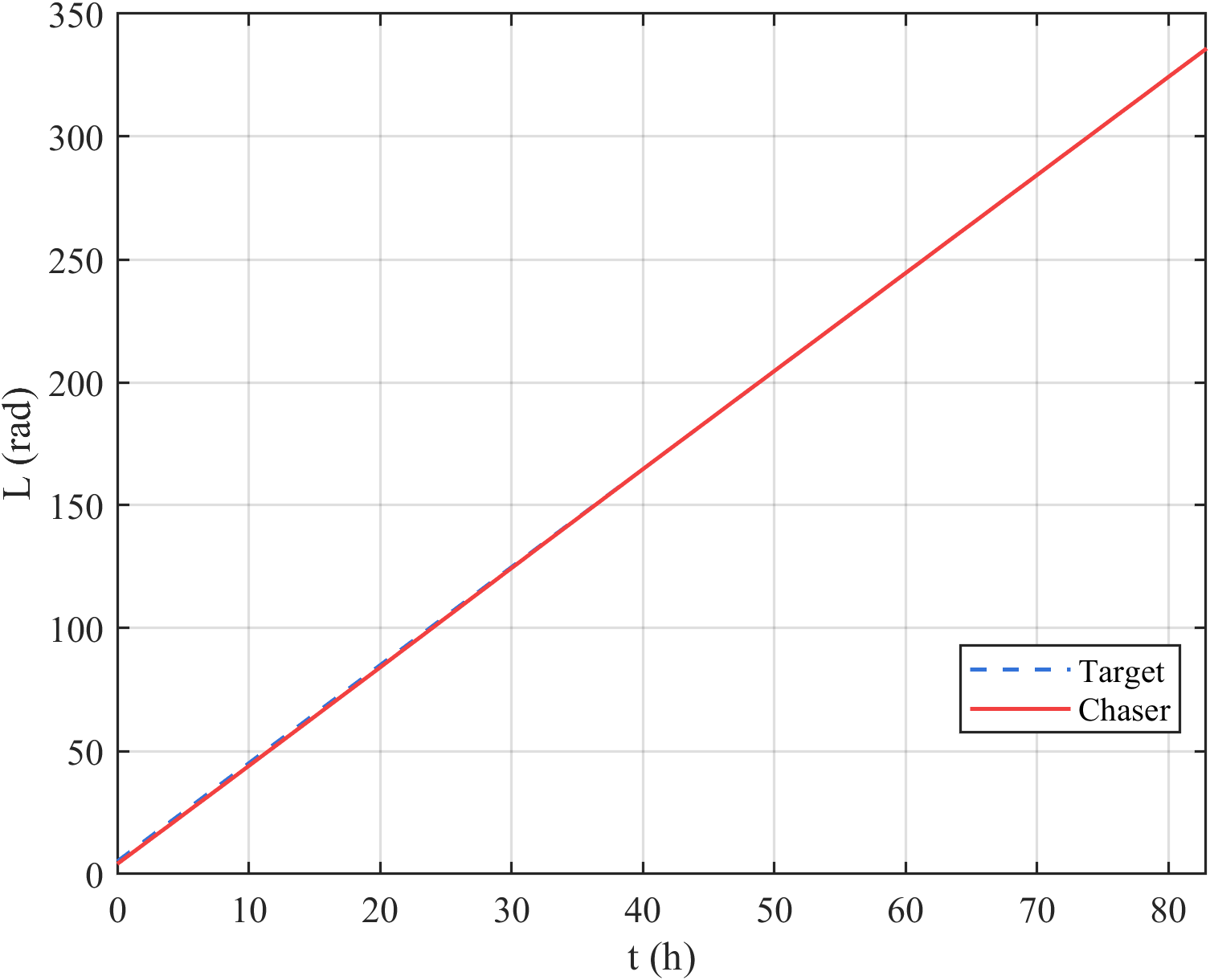}
        \label{fig:mee_L}
    }
    \caption{Evolution of MEE.}
    \label{fig:mee_history}
\end{figure}

The performance metrics of the methods evaluated  for this specific maneuver are summarized in Table \ref{tab:comparison_results}. The metrics include the time of flight (ToF), terminal position error (Pos. Error),  terminal velocity error (Vel. Error),and fuel consumption ($\Delta m$). The trajectory of the chaser and target is depicted in Fig.\ref{fig:trajectory_comp}, while Fig.\ref{fig:mee_history} illustrates the temporal variations of the MEEs.

\begin{table}[htbp]
\centering
\caption{Performance comparison of the evaluated strategies for different test cases.}
\label{tab:comparison_results}
\begin{tabular}{@{}lcccc@{}}
\toprule
\textbf{Method} & \textbf{ToF} (h) & \textbf{Pos. Error} (m) & \textbf{Vel. Error} (m/s) & \textbf{Fuel Cons.} ($\Delta m$, kg) \\ \midrule
\multicolumn{5}{l}{\textit{Case 1: Representative Case}} \\
Proposed Method & 79.78 & 140.88 & $0.414$ & 0.49 \\
Direct Method   & 120.1784 & $4.53\times10^{-7}$ & $6.61\times 10^{-11}$ & 0.32 \\
Hybrid Method   & 82.46 & 0.44 & $3.78\times 10^{-4}$ & 0.16 \\ \midrule
\multicolumn{5}{l}{\textit{Case 2: Long-Horizon Case}} \\
Proposed Method & 231.14 & 163.93 & 0.92 & 1.33 \\
Direct Method   & 196.89 & $1.74\times 10^{-3}$ & $6.28\times 10^{-6}$ & 1.48 \\
Hybrid Method   & 231.44 & 0.28 & $2.82\times10^{-4}$ & 1.29 \\ \bottomrule
\end{tabular}
\end{table}

The results demonstrate that the proposed method can successfully accomplish multi-revolution rendezvous missions across both representative and long-horizon scenarios. Although the learned policy does not explicitly optimize a specific performance metric such as propellant consumption or time of flight, it is capable of generating feasible rendezvous trajectories with competitive performance. As shown in Table \ref{tab:comparison_results}, the proposed method exhibits advantages in either time of flight or propellant consumption depending on the mission scenario, highlighting the effectiveness of the learned policy in balancing multiple objectives across a broad range of rendezvous conditions.

{Residual terminal errors remain under the present policy and control-update configuration. The available experiments do not isolate their cause among policy approximation, training objectives, and control discretization.} For example, in Case 1, the final position error is approximately $140.88\,\mathrm{m}$. Therefore, further improving terminal rendezvous accuracy remains an important direction for future research. In addition, the hybrid method effectively bridges this precision gap. By providing a high-quality initial guess for the subsequent optimization process, the learned policy significantly improves the convergence behavior of the numerical solver and enables high-precision rendezvous solutions for challenging long-duration missions.

In conclusion, the proposed method successfully accomplished the multi-revolution rendezvous with a suboptimal and real-time control law. Furthermore, it robustly ensures the convergence of NLP solvers for high-precision spaceflight operations when employed as an initial guess generator.

\subsection{Ablation Studies}
\label{subsec:ablation}

\begin{figure}[htbp]
    \centering
    \subfloat[Reward\label{fig:reward}]{
        \includegraphics[width=0.48\linewidth]{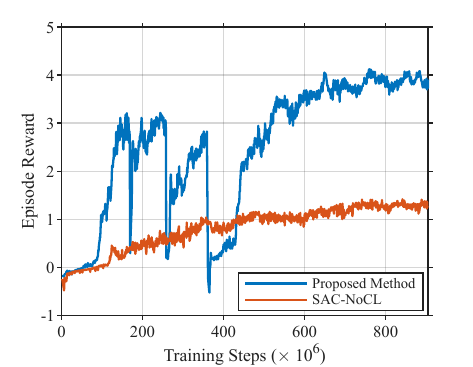}
    }\hfill
    \subfloat[Q-Value\label{fig:q_value}]{
        \includegraphics[width=0.48\linewidth]{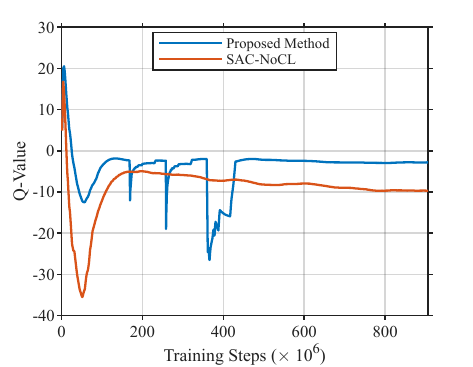}
    }\hfill
    \subfloat[Actor Loss\label{fig:actor_loss}]{
        \includegraphics[width=0.48\linewidth]{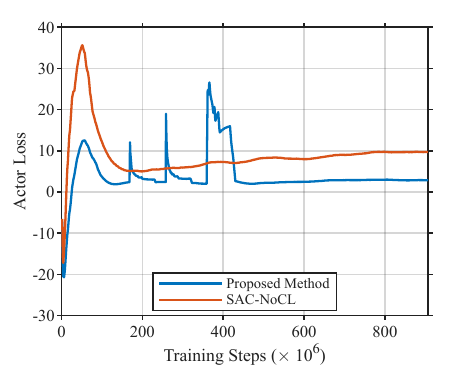}
    }\hfill
    \subfloat[Success Rate\label{fig:success_rate}]{
        \includegraphics[width=0.48\linewidth]{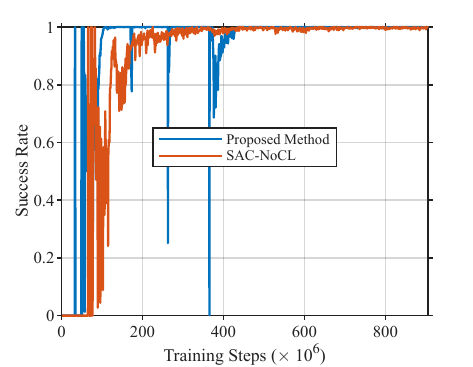}
    }
    \caption{Comparative evolution of the training metrics}
    \label{fig:losses}
\end{figure}

An ablation study is conducted to demonstrate the necessity of the CL strategy. The progression of the curriculum learning strategy for the Proposed Method is quantitatively summarized in Table \ref{tab:curriculum_steps}. The success rate threshold for advancing to the next difficulty level is strictly set at 90\%. In this study, the evolution of the training metrics for both the proposed method and the SAC-NoCL method, evaluated during the interaction steps $8.5 \times 10^8$, is presented in Fig.\ref{fig:losses}.

\begin{table}[htbp]
\centering
\caption{Cumulative training steps and durations at curriculum stage transitions.}
\label{tab:curriculum_steps}
\begin{tabular}{@{}lcc@{}}
\toprule
\textbf{Curriculum Transition} & \textbf{Cumulative Training Steps} & \textbf{Duration in Current Stage} \\ \midrule
Stage 1 $\rightarrow$ Stage 2 & ${168,320,000}$ & ${168,320,000}$ \\
Stage 2 $\rightarrow$ Stage 3 & ${257,920,000}$ & ${89,600,000}$ \\
Stage 3 $\rightarrow$ Stage 4 & ${360,320,000}$ & ${102,400,000}$ \\ \bottomrule
\end{tabular}
\end{table}

As depicted in Fig.\ref{fig:losses} , the removal of the CL strategy does not drastically compromise the convergence of the training process. Both the actor loss and Q-value  of the SAC-NoCL agent stabilize, indicating that the baseline agent is effectively updating its parameters to minimize the Bellman error within its explored state distribution. 

However, as shown in Fig.\ref{fig:reward} and Fig.\ref{fig:q_value}, the SAC-NoCL agent plateaus with a significantly lower reward and lower Q-value. In addition, the agent trained using the SAC-NoCL method did not discover the optimal strategies parameterized by the proposed curriculum learning framework. To quantify the degradation in policy optimality, Table \ref{tab:ablation_results} presents a direct comparison of mission performance between the two methods in the case defined in Table \ref{tab:test_cases} case 1.

\begin{table}[htbp]
\centering
\caption{Ablation comparison of mission optimality.}
\label{tab:ablation_results}
\begin{tabular}{@{}lcccc@{}}
\toprule
\textbf{Method} &
\textbf{ToF} (h) &
\textbf{Pos. Error} (m) &
\textbf{Vel. Error} (m/s) &
\textbf{Prop. Cons.} ($\Delta m$, kg) \\
\midrule

Proposed Method
& {79.78}
& {140.88}
& {$0.414$}
& {0.49} \\

SAC-NoCL
& {337.33}
& {450.25}
&{2.62}
& {2.12} \\

\bottomrule
\end{tabular}
\end{table}

\begin{figure}[htbp]
    \centering
    \subfloat[Element $p$ (km)]{
        \includegraphics[width=0.32\linewidth]{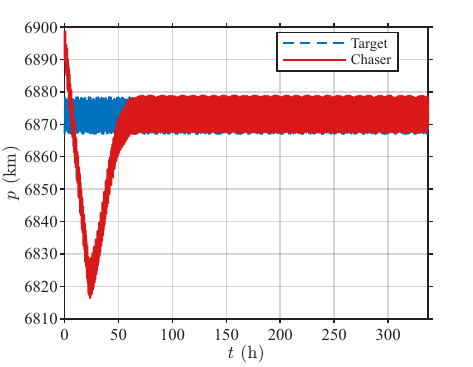}
        \label{fig:nocl_p}
    }\hfill
    \subfloat[Element $f$]{
        \includegraphics[width=0.32\linewidth]{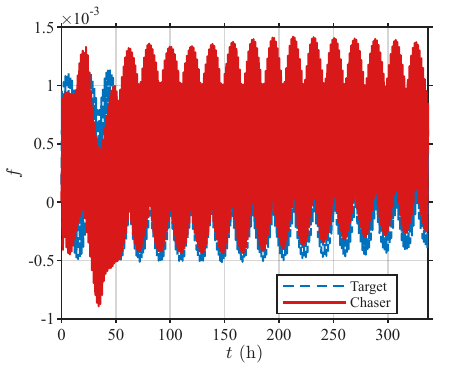}
        \label{fig:nocl_f}
    }\hfill
    \subfloat[Element $g$]{
        \includegraphics[width=0.32\linewidth]{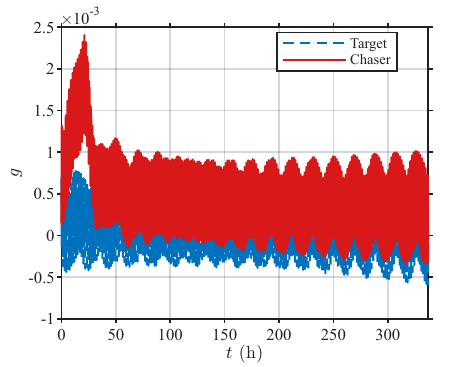}
        \label{fig:nocl_g}
    }\hfill
    \subfloat[Element $h$]{
        \includegraphics[width=0.32\linewidth]{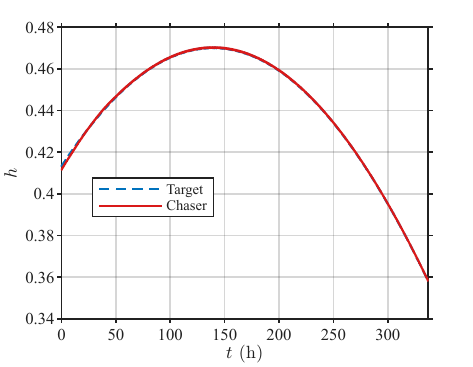}
        \label{fig:nocl_h}
    }\hfill
    \subfloat[Element $k$]{
        \includegraphics[width=0.32\linewidth]{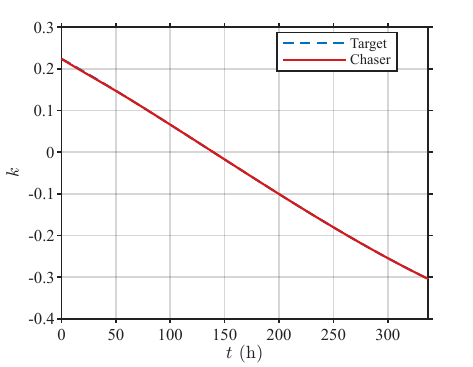}
        \label{fig:nocl_k}
    }\hfill
    \subfloat[True Longitude $L$ (rad)]{
        \includegraphics[width=0.32\linewidth]{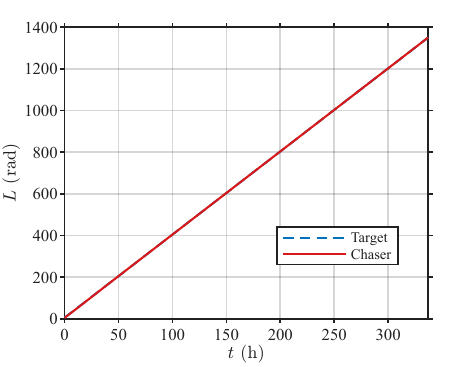}
        \label{fig:nocl_L}
    }
    \caption{Time-history evolution of MEEs for the SAC-NoCL baseline.}
    \label{fig:nocl_mee}
\end{figure}

The empirical results reveal a pronounced performance degradation for the ablation baseline. Compared with the proposed method, the SAC-NoCL approach requires approximately $4$ to $5$ times longer flight times and exhibits noticeably larger terminal position and velocity errors. The underlying reason can be observed from the temporal evolution of the MEEs in Fig.~\ref{fig:nocl_mee}. In contrast to the proposed method shown in Fig.~\ref{fig:mee_history}, the SAC-NoCL trajectory evolves over a much longer time horizon and maintains a persistent offset from the target trajectory for most of the transfer. The periodic variations observed in $p$, $f$, and $g$ are mainly induced by orbital motion rather than by control oscillations. Nevertheless, the delayed convergence of the MEE components, together with the prolonged evolution of the true longitude $L$, indicates that the policy without curriculum learning fails to guide the chaser toward the target efficiently. This explains the increased flight time and degraded terminal accuracy of the SAC-NoCL baseline.

The result of the ablation studies demonstrates that, without the progressive structural constraints imposed by curriculum learning, the DRL agent fails to decouple the multi-revolution rendezvous requirement from short-term orbital shape preservation. Consequently, the agent resorts to persistent trial-and-error maneuvering instead of executing a one-shot intercept. The result validates that, while the maximum-entropy framework of SAC ensures convergence, CL is indispensable for guiding the policy toward efficient and optimal regions in complex rendezvous tasks.

\subsection{Computational Efficiency}
\label{subsec:efficiency}

To quantitatively evaluate the computational acceleration achieved by the proposed fully tensorized JAX architecture, a comprehensive performance benchmark was conducted. It is essential to isolate the speedup strictly attributable to the hardware-software co-design from other algorithmic factors. Therefore, we first explicitly define the hardware platform and the baseline used for this comparative analysis.

\paragraph{Hardware setup and software stack}

All simulations, environmental vectorization, and neural network training in this study were executed on a dedicated high-performance compute node. The hardware platform is equipped with an AMD EPYC 9654 central processing unit (CPU) and a single NVIDIA L40 graphics processing unit (GPU) with 48 GB of GDDR6 VRAM. The software stack is built upon Ubuntu 22.04.5 LTS, Python 3.10, and JAX version 0.6.2, utilizing CUDA 12.4 and cuDNN toolkits to accelerate tensor operations.

Crucially, given the extreme scale disparities inherent in astrodynamics—ranging from the Earth's radius ($\sim 10^7$ m) to low-thrust accelerations ($\sim 10^{-4}$ $\text{m/s}^2$)—the JAX framework was explicitly configured to enable double-precision floating-point operations (FP64, \texttt{jax\_enable\_x64=True}). This strict 64-bit precision guarantees the numerical stability and physical fidelity of the 5th-order Runge-Kutta integration during long-horizon rollouts. {In the training pipeline, both nominal $J_2$ orbital propagation and SAC policy optimization are executed on the same GPU.}

\paragraph{Benchmark baseline definition}

To establish a perfectly controlled baseline and ensure algorithmic parity, the conventional CPU baseline was constructed using the exact same JAX-based codebase. Instead of relying on a disparate software library (which could introduce algorithmic confounding factors), the identical tensorized environment—incorporating the 5th-order Runge-Kutta integrator and $J_2$ perturbation models—was compiled via XLA and explicitly dispatched to the CPU backend.

To represent a standard high-end workstation and to circumvent the severe Non-Uniform Memory Access penalties associated with multi-socket servers, the XLA CPU thread pool was strictly constrained to utilize an allocation of 16 physical CPU cores.

In a conventional heterogeneous reinforcement learning setup, deploying the environment on the CPU while hosting the neural networks on the GPU fundamentally necessitates continuous data transfers (i.e., observations, rewards, and actions) via the PCIe bus. By executing the identically coded environment on both the CPU baseline and the proposed GPU architecture, we perfectly isolate the hardware-level performance gains. This setup rigorously demonstrates how migrating the entire computational graph to a unified GPU-resident memory space—combined with the immense throughput of SIMT processing—overcomes both the CPU compute limitations and the PCIe communication bottlenecks.

\subsubsection{Simulation Throughput Analysis}
\label{subsubsec:scalability}

To evaluate the raw computational efficiency of the physics engine, an isolated throughput benchmark was conducted, temporarily decoupling the SAC neural network optimization. The metric utilized is Steps Per Second (SPS), representing the total number of orbital integration steps executed across all parallel environments in one second of wall-clock time. The agent was configured to output random control actions to drive the environmental state transitions.

\begin{figure*}[t]
\centering
\includegraphics[width=0.8\linewidth]{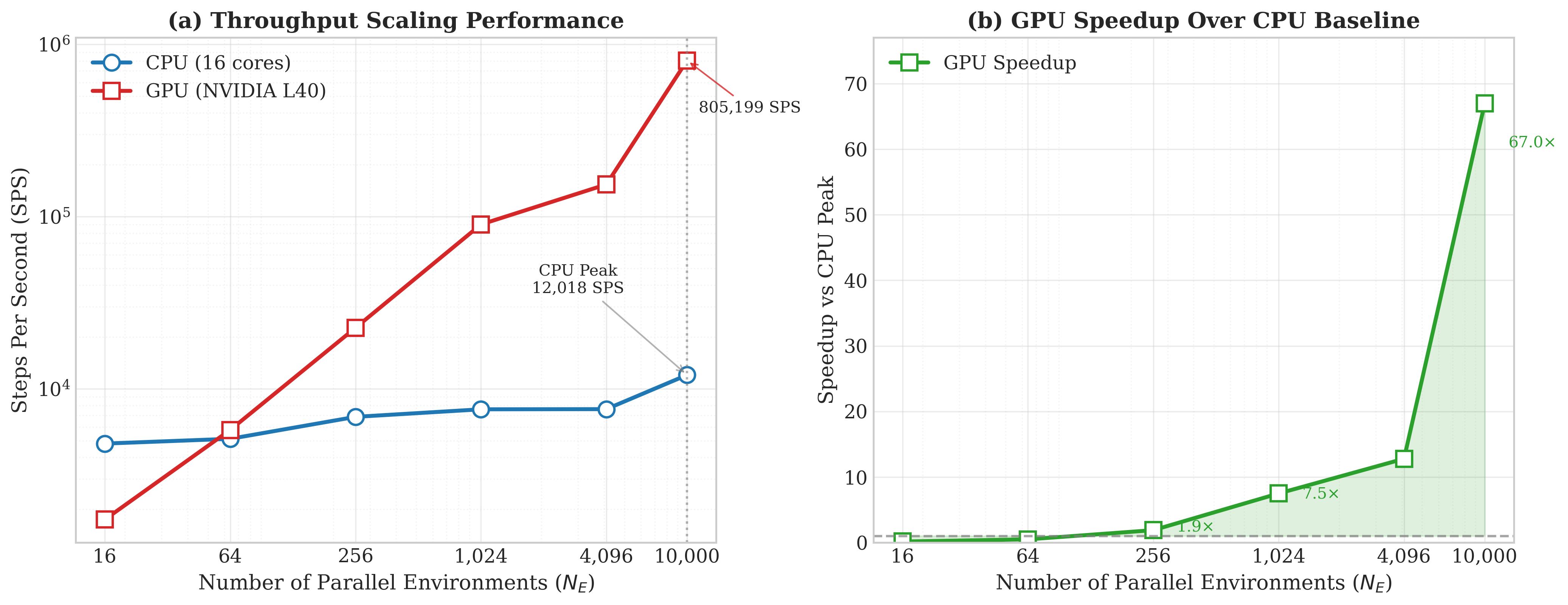}
\caption{Computational throughput and scalability analysis. (a) Steps Per Second (SPS) achieved by the 16-core CPU baseline and the NVIDIA L40 GPU as a function of the number of parallel environments ($N_E$), plotted on a logarithmic scale. (b) The corresponding speedup ratio of the GPU architecture relative to the CPU peak performance, reaching a maximum of 67.0$\times$ at $N_E = 10000$.}
\label{fig:throughput}
\end{figure*}

Fig.\ref{fig:throughput} illustrates the SPS scaling performance of both architectures as the number of parallel environments ($N_E$) increases from 16 to 10,000. For the CPU baseline, a resource allocation of 16 physical cores was deliberately selected to represent a standard high-performance workstation configuration, while avoiding the severe Non-Uniform Memory Access overheads and thread contention typical of highly concurrent XLA workloads on multi-socket CPUs. As depicted, the CPU baseline exhibits near-linear scaling initially but sharply saturates and subsequently degrades once $N_E$ exceeds the allocated core count (16 cores), peaking at only 12,018 SPS. This degradation is primarily attributed to the prohibitive overhead of context switching and cache thrashing when oversubscribing CPU threads.

\subsubsection{End-to-End Training Performance}
\label{subsubsec:end_to_end}

Conversely, the JAX-based tensorized architecture demonstrates exceptional scalability. It is worth noting that at extremely low concurrency (e.g., $N_E \le 64$), the CPU marginally outperforms the GPU. This is a well-known characteristic of heterogeneous computing, where the fixed overhead of GPU kernel launching and XLA dispatching dominates the execution time for small workloads. However, as $N_E$ increases, the GPU's advantage becomes overwhelmingly apparent. By leveraging the \texttt{vmap} operator and XLA compilation, the orbital dynamics are strictly vectorized, allowing the NVIDIA L40 GPU to fully unleash its SIMT capabilities. At the massively parallel scale of $N_E = 10000$, the JAX environment achieves a staggering throughput of 805,199 SPS, realizing a maximum speedup of 67.0$\times$ compared to the peak performance of the 16-core CPU baseline.

While the isolated throughput metric demonstrates the raw efficiency of the vectorized physics engine, the ultimate metric for deep reinforcement learning is the end-to-end wall-clock time required for policy convergence. In this subsection, we profile the complete SAC training pipeline, which simultaneously encompasses environment rollouts, replay buffer management, and neural network backpropagation. To rigorously isolate the communication bottlenecks inherent in traditional frameworks, we conduct a three-architecture ablation study evaluated at a moderate concurrency scale ($N_E = 320$) for $10^5$ transitions. Crucially, to ensure strict algorithmic parity across all hardware configurations, the workload is constrained to absolute equivalence: for every block of 10,000 environment transitions collected, exactly 10 SAC gradient updates are executed.

In the conventional heterogeneous baseline (CPU for environment rollouts, GPU for neural network training), the training loop is severely bottlenecked by data fragmentation. At every interaction step, massive arrays of observations and rewards must be serialized and transferred from the host memory (CPU) to the device memory (GPU), while the corresponding continuous action tensors are transmitted back via the PCIe bus. As shown in Table \ref{tab:end_to_end_time}, while this hybrid approach accelerates the pure CPU baseline by 17.9$\times$ (achieving 6,616 SPS), this recurrent communication overhead paralyzes the GPU's computational units, forcing the Tensor Cores to remain idle while awaiting data synchronization.

The proposed fully tensorized JAX-based architecture explicitly resolves this bottleneck by unifying the environment and the policy network within the same computational graph. Through a "zero-copy" execution flow, the vectorized state transitions generated by the RK5 integrators are directly ingested into the SAC Replay Buffer, which resides entirely within the GPU VRAM.

Quantitative benchmarking confirms the superiority of this unified approach. By minimizing rollout-related host-to-device communication overhead, the GPU-rollout and GPU-training architecture achieves 10,389 SPS, representing an approximately 28.2$\times$ speedup over the pure 16-core CPU baseline and a 57\% throughput increase over the heterogeneous hybrid architecture. When extrapolated to a full training run of $8.5 \times 10^8$ transitions, the proposed framework reduces the computational time from approximately 640 hours on the pure CPU baseline to about 22.7 hours. This substantial reduction in wall-clock time makes large-scale policy training practically feasible.

\begin{table*}[htbp]
\centering
\caption{End-to-end performance ablation for collecting and training on $10^5$ transitions ($N_E = 320$).}
\label{tab:end_to_end_time}
\begin{tabular}{@{}lcccc@{}}
\toprule
\textbf{Architecture Setup} & \textbf{Rollout Backend} & \textbf{Training Backend} & \textbf{End-to-End SPS} & \textbf{Relative Speedup} \\ \midrule
Pure CPU (Baseline) & CPU (16-Cores) & CPU (16-Cores) & 369 & 1.0$\times$ \\
Heterogeneous Hybrid & CPU (16-Cores) & GPU (L40) & 6,616 & 17.9$\times$ \\
\textbf{Fully Tensorized (Proposed)} & \textbf{GPU (L40)} & \textbf{GPU (L40)} & \textbf{10,389} & \textbf{28.2$\times$} \\ \bottomrule
\end{tabular}
\end{table*}

\subsubsection{Real-Time Inference Efficiency}

Beyond offline training acceleration, the practical viability of a guidance algorithm is fundamentally dependent on its online execution efficiency. Table \ref{tab:inference_time} compares the computational time required to generate the full multi-revolution trajectory.

The proposed method achieves an order-of-magnitude speedup over the direct method, This efficiency arises because RL inference relies purely on lightweight neural network forward propagation and RK5 integration, fundamentally bypassing the computationally expensive numerical iterations and Jacobian evaluations inherent in NLP solvers. Although the sequential hybrid method naturally requires the longest computation time, it ensures extreme terminal precision for robust offline trajectory generation. Ultimately, the framework successfully shifts the heavy optimization burden to the offline training stage, producing a highly efficient inference process capable of enabling real-time onboard closed-loop guidance under limited computational resources.

\begin{table}[htbp]
\centering
\caption{Computational time comparison during the inference phase for different cases.}
\label{tab:inference_time}
\begin{tabular}{@{}llc@{}}
\toprule
\textbf{{Case}} & \textbf{Method} & \textbf{Comp. Time (s)} \\ \midrule
\multirow{3}{*}{{Case 1}} & Proposed Method & {32.77} \\
                        & Direct Method   & {510.30} \\
                        & Hybrid Method   & {1087.50} \\ \midrule
\multirow{3}{*}{{Case 2}} & Proposed Method & {56.35}  \\ 
                        & Direct Method   & {515.20}  \\ 
                        & Hybrid Method   & {1090.10} \\ \midrule
\end{tabular}
\end{table}

\section{Conclusion}
\label{sec:conclusion}

This paper proposes a massively parallel deep reinforcement learning framework to address the computational and algorithmic challenges of long-horizon, continuous-thrust multi-revolution rendezvous in low Earth orbit. By integrating a JAX-based tensorized astrodynamic environment with the soft actor-critic algorithm and a progressive curriculum learning strategy, the proposed approach successfully shifts the heavy optimization burden from online computation to offline training.

Computationally, the fully GPU-resident, zero-copy architecture eliminates the PCIe communication bottlenecks typical of heterogeneous reinforcement learning setups, achieving a 67-fold increase in simulation throughput compared to conventional multi-core CPU baselines. Algorithmically, this extreme data throughput, coupled with the progressive structural constraints of the four-stage curriculum, effectively overcomes the sparse reward and long-term credit assignment issues inherent in $J_2$-perturbed orbital dynamics. Extensive simulations validate that the curriculum-enhanced agent robustly synthesizes suboptimal guidance policies across a broad domain of initial conditions, avoiding the severe performance collapse observed in standard off-policy learning.

Crucially, the practical viability of the learned policy is demonstrated through its highly efficient online execution. Using purely neural network forward propagation and numerical integration, the proposed method accelerates trajectory generation by an order of magnitude compared to traditional direct collocation methods, completing complex phasing maneuvers in seconds. Furthermore, the RL policy serves as a highly reliable warm-start generator for NLP solvers. This hybrid pipeline bridges the precision gap of continuous-action policy approximations by using the RL-generated trajectory as an informed initial guess for the subsequent NLP refinement. Compared with uninformed or randomly initialized guesses, this warm-start strategy improves the convergence behavior of the NLP solver.

{High-fidelity closed-loop evaluations show that the frozen policy maintains high rendezvous success rates under the tested perturbations, eclipse-induced thrust interruptions, and observation errors. These results support robustness within the evaluated conditions, but do not establish meter-level accuracy.}

Ultimately, this hardware-software co-design paradigm provides a scalable and computationally lightweight solution for closed-loop orbital guidance. The demonstrated balance of real-time inference capability, algorithmic robustness, and physical fidelity positions this framework as a practical enabler for autonomous constellation management and on-orbit servicing missions under limited onboard computational resources.

 \bibliographystyle{elsarticle-num} 
 \bibliography{references}

\section*{Declaration of generative AI and AI-assisted technologies in the manuscript preparation process}

During the preparation of this work the author(s) used Gemini (Google) {and OpenAI Codex} in order to improve the language, readability, and organization of the manuscript, as well as to optimize the JAX-based implementation logic and figure presentation strategies. After using this tool, the author(s) reviewed and edited the content as needed and take(s) full responsibility for the content of the published article.
\end{document}